\documentclass[%
 preprint,
nofootinbib,
 amsmath,amssymb,
 aps,
]{revtex4-2}

\usepackage{graphicx}
\usepackage{dcolumn}
\usepackage{bm}
\usepackage{hyperref}
\usepackage[mathlines]{lineno}

\usepackage{xcolor}

\begin{document}

\preprint{}

\title{Not all that is $\beta_0$ is $\beta$-function: the DGLAP resummation and the running coupling in NLO JIMWLK }

\author{Alex Kovner}\email{alexander.kovner@uconn.edu}

\affiliation{Physics Department, University of Connecticut, 196A Auditorium road, Storrs, CT 06269-3046, USA}

\author{Michael Lublinsky}
\affiliation{Physics Department, Ben-Gurion University of the Negev, Beer Sheva 84105, Israel}

\author{Vladimir~V.~Skokov}
 \email{vskokov@ncsu.edu}

\affiliation{Department of Physics, North Carolina State University, Raleigh, NC 27695, USA}

\affiliation{RIKEN/BNL Research Center, Brookhaven National
  Laboratory, Upton, NY 11973, USA}

\author{Zichen Zhao}
\affiliation{Department of Physics, North Carolina State University, Raleigh, NC 27695, USA}

\date{\today}

\begin{abstract}
We reanalyze the origin of the large transverse logarithms associated with the QCD one loop $\beta$ function coefficient in the NLO JIMWLK Hamiltonian. We show that some of these terms are not associated with the running of the QCD coupling constant but rather with the DGLAP evolution. 
The DGLAP-like resummation of these logarithms is mandatory within the JIMWLK Hamiltonian, as long as the color correlation length in the projectile is larger than that in the target. This regime in fact  covers the whole range of rapidities at which JIMWLK evolution is supposed to be applicable. 
We derive the RG equation that resums these logarithms to all orders in $\alpha_s$ in the JIMWLK Hamiltonian. This is a nonlinear equation for the eikonal scattering matrix $S(\mathbf{x)}$. We solve this equation, and perform the DGLAP resummation in two simple cases: the dilute limit, where both the projectile and the target are far from saturation, and the saturated regime, where the target correlation length also determines its saturation momentum. 

\end{abstract}

\maketitle

\tableofcontents
\newpage 

\section{Introduction}
At high energy, the energy evolution of hadronic cross sections (see e.g. 
\cite{Gelis:2010nm} for review) is governed by the the Jalilian-Marian, Iancu, McLerran, Weigert, Leonidov and Kovner (JIMWLK) evolution equation~\cite{Jalilian-Marian:1997qno,Jalilian-Marian:1997ubg,Kovner:2000pt,Iancu:2000hn,Ferreiro:2001qy}. The JIMWLK equation was derived at next-to-leading order (NLO) almost ten years ago~\cite{Kovner:2013ona,Kovner:2014lca,Lublinsky:2016meo}, following the earlier derivation of the NLO dipole model (Balitsky-Kovchegov equation), see~\cite{Balitsky:1995ub,Balitsky:1997mk,Kovchegov:1999yj,Balitsky:2007feb}. With the derivation, it became immediately apparent that improvements are necessary due to the presence 
of large transverse logarithms in the NLO kernel. Two types of terms contain large transverse logarithms in the NLO JIMWLK equation. The first type is proportional to the one loop coefficient of the QCD $\beta$-function, $\beta_0$; while the second does not involve the $\beta$-function and is related to the lifetime ordering of fluctuations.

Different recipes for the resummation of both types of logarithms have been extensively discussed in the literature in recent years~
\cite{Kutak:2004ym,Motyka:2009gi,Beuf:2014uia,Iancu:2015vea,Iancu:2015joa,Vera:2005jt,Ducloue:2019ezk}.
In this paper, we will have nothing to say about the second type of logarithms, which we will ignore  from now on, with the realization that the physics that goes into their resummation is very different from the one discussed here.
The consensus for the $\beta_0$-dependent logarithms has been that they should all be resummed into the running of the QCD coupling constant~\cite{Balitsky:2006wa,Kovchegov:2006vj}, see also 
\cite{Gardi:2006rp}. Different recipes have been advanced for the optimal scale setting in the running coupling~\footnote{As was investigated in Ref.~\cite{Altinoluk:2023}, these schemes violate the semi-positivity of the JIMWLK Hamiltonian.}. 

In this paper, we revisit the resummation of this type of logarithms. We show that not all of them are associated with the running of the coupling, but that the correct understanding of a subset of them is in terms of the {\it DGLAP evolution}~\cite{Dokshitzer:1977sg,Gribov:1972ri,Altarelli:1977zs}  of the projectile wave function. We explicitly set up the (transverse) evolution equation for the JIMWLK Hamiltonian 
that resums these logarithms into the modified JIMWLK kernel. We also provide an approximate solution to this equation in two simple cases: the dilute limit and the limit of dense saturated target. The main qualitative result of this resummation is the smearing of the emission point of the gluon in the JIMWLK Hamiltonian over the transverse resolution scale with a function decreasing as a power of a distance. Physically this transverse resolution scale is provided by the saturation momentum of the color fields in the target. Our analysis also suggests a simple scale setting procedure for the QCD running coupling, which does not require any elaborate considerations. Together with the DGLAP resummation, this simple scale setting procedure takes care of all the large transverse logarithms\footnote{Again,  in this statement, we do not include the logarithms associated with the lifetime ordering of fluctuations.}

Our discussion is set explicitly in the context of a ``dense'' target and ``dilute'' projectile in the following sense. We assume that the color fields in the target are correlated on a transverse distance scale $Q_T^{-1}$. In the projectile we also assume the existence of an analogous scale $Q_P^{-1}$, and we  require that $Q_T\gg Q_P$, i.e. that the correlation length of the color fields in the target is much smaller than that in the projectile.

The existence of such correlation length is natural in the saturation regime. In this case we identify $Q_T$ with the saturation momentum in the target, and analogously for the projectile. The hierarchy of the correlation length then just means that the target has a larger saturation momentum. However our approach is also applicable when neither target nor projectile are saturated, i.e. the eikonal scattering factors are perturbatively close to unity. In this case the scale $Q^{-1}$ has the meaning of the transverse scale on which the color neutralization occurs in the wave function. For example, for a single dipole projectile, $Q_P^{-1}$ would be of the order of the dipole size. In this case the hierarchy of scales means that the target ``dipole'' has smaller transverse size than the projectile ``dipole'', so that it is still true that the projectile is more dilute than the target. Indeed only in the case when one has such large separation of scales, the DGLAP resummation is required.

This paper is structured as follows. In Sec.~\ref{Sec2} we review the NLO JIMWLK Hamiltonian and the common lore about the resummation of $\beta_0$ dependent transverse logarithms. In Sec.~\ref{Sec3} we introduce the physical interpretation in terms of transverse resolution scale and formalize it by defining (in the JIMLWK framework) the analog of factorization scale in DIS. We derive the RG equation for the JIMWLK Hamiltonian which imposes the overall factorization scale independence of physical amplitudes, and discuss what is the best choice for initial and final scales in this evolution. In Sec.~\ref{Sec4} we provide an approximate solution for this equation and discuss its properties in the absence of quarks. In Sec.\ref{Sec5} we generalize the discussion to include quarks. Finally, we conclude in Sec.~\ref{Sec6} with a discussion of our results.

\section{JIMWLK at NLO and the large transverse logarithms}
\label{Sec2}
The JIMWLK Hamiltonian is the operator that generates rapidity evolution of the hadronic $S$-matrix at high energy.  We take it here as acting on the projectile wave function (see Ref.~\cite{Kovner:2005pe} for a comprehensive introduction), or rather probability distribution $\mathcal{W}_P$, so that the rapidity evolution of the scattering matrix of a hadronic projectile on a hadronic target is given by
\begin{equation}\label{evolution}
\frac{d}{dY}{\cal S}=\langle H_{\rm JIMWLK}[S,J]\mathcal W_P[S]\rangle_T
\end{equation}
where $\cal S$ is the hadronic $S$-matrix, $S$-is an eikonal propagation factor (Wilson line) of a projectile parton on a target field, $J$ is a rotation operator (see below) and the averaging on RHS is performed over the ensemble of the target fields.
 
\subsection{JIMWLK at Leading Order}
At leading order the JIMWLK Hamiltonian (see e.g. Ref.~\cite{Lublinsky:2016meo}) is given by
 \begin{equation}H^{\rm LO}_{\rm JIMWLK}
  =\,\frac{\alpha_{s}}{2\pi^{2}}\int_{\mathbf{x},\, \mathbf{y},\, \mathbf{z}}\frac{X\cdot Y}{X^{2}Y^{2}}\,\left[\, J_{L}^{a}(\mathbf{x})\, J_{L}^{a}(\mathbf{y})\,+\, J_{R}^{a}(\mathbf{x})\, J_{R}^{a}(\mathbf{y})\,-\,2J_{L}^{a}(\mathbf{x})\, S^{ab}(\mathbf{z})\, J_{R}^{b}(\mathbf{y})\,\right].
  \end{equation}
  Here $X_i\equiv \mathbf{x}_i-\mathbf{z}_i$ and $Y_i\equiv \mathbf{y}_i-\mathbf{z}_i$. 
  The LO kernel is a product of two Weizsacker-Williams (WW) fields. 
  $S$ is the gluon's eikonal scattering matrix, that is the Wilson line in adjoint representation.
 The left and right rotation operators $J^a_L(\mathbf{x})$ and $J^a_R(\mathbf{x})$ generate left and right rotations of the Wilson line $S(\mathbf{x})$ at the same transverse point $\mathbf{x}$. 
 \begin{equation}
J^a_R(\mathbf{x}){S}^{eb}(\mathbf {y})=[S(\mathbf{y})T^a]^{eb}\delta^2(\mathbf{x}-\mathbf{y});\ \ \ \ 
J^a_R(\mathbf{x}){S}^{T eb}(\mathbf {y})=-[T^aS^T(\mathbf{y})]^{eb}\delta^2(\mathbf{x}-\mathbf{y})\,.
\end{equation}
$J_R$ and $J_L$  are related by 
 \begin{equation}\label{rl}
 J^a_R(\mathbf{x})=S^{\dagger ab}(\mathbf{x})J^b_L(\mathbf{x})\,.
 \end{equation}

 \subsection{JIMWLK Hamiltonian at NLO - the relevant terms}
 \label{sdfla}
The NLO JIMWLK Hamiltonian deduced in Refs.~\cite{Kovner:2013ona,Kovner:2014lca} is:
\begin{eqnarray}\label{ham}
&&H_{\rm JIMWLK}^{\rm NLO}=\int_{\mathbf{x}, \mathbf{y}, \mathbf{z}}K_{JSJ}(\mathbf{x}, \mathbf{y},\mathbf{z})\left[J_{L}^{a}(\mathbf{x}) J_{L}^{a}(\mathbf{y})+J_{R}^{a}(\mathbf{x})J_{R}^{a}(\mathbf{y})-2J_{L}^{a}(\mathbf{x}) S^{ab}(\mathbf{z})J_{R}^{b}(\mathbf{y})\right]\ \nonumber\\
&&+\int_{\mathbf{x}, \mathbf{y},\mathbf{z}, \mathbf{z}^{\prime}} K_{JSSJ}(\mathbf{x},\mathbf{y},\mathbf{z}, \mathbf{z}^{\prime})\left[J_{L}^{a}(\mathbf{x})D^{ad}(\mathbf{z},\mathbf{z}^{\prime}) J_{R}^{d}(\mathbf{y})-\frac{N_c}{2}J_R^a(\mathbf{x})J_R^a(\mathbf{y})-\frac{N_c}{2}J_L^a(\mathbf{x})J_L^a(\mathbf{y})\right]
\\
&&+\int_{\mathbf{x}, \mathbf{y}, \mathbf{z},\mathbf{z}^{\prime}}K_{q\bar{q}}(\mathbf{x}, \mathbf{y},\mathbf{z}, \mathbf{z}^{\prime})\left[2\, J_{L}^{a}(\mathbf{x})\, tr[V^{\dagger}(\mathbf{z})\, t^{a}\, V(\mathbf{z}^{\prime})t^{b}]\, J_{R}^{b}(\mathbf{y})-\frac{1}{2}J_R^a(\mathbf{x})J_R^a(\mathbf{y})-\frac{1}{2}J_L^a(\mathbf{x})J_L^a(\mathbf{y})\right]\,
\nonumber\\
&&+...\nonumber 
\end{eqnarray}
where, to simplify notations, we introduced 
\begin{align}
    D^{ab}(\mathbf{z}_1, \mathbf{z}_2) = 
    {\rm Tr}[ T^a  S(\mathbf{z}_1)  T^b S^+(\mathbf{z}_2)]\,
\end{align}
while ellipsis denotes terms that do not depend on $\beta_0$ and do not play any role in our discussion.
 In Eq.~\eqref{ham},  the rotation operators $J_{L}$ and $J_{R}$ do not act on the eikonal factors $S$ in the Hamiltonian itself, but only on those in the wave function ${\mathcal W}_P[S]$ in Eq.~\eqref{evolution}. 

The kernels $K$ in this equation are given by the following expressions.
\begin{equation}\begin{split} \label{kprime}
K_{JSJ}(\mathbf{x},\mathbf{y},\mathbf{z})&\equiv\,\frac{\alpha_{s}^{2}(\mu)X\cdot Y}{4\pi^{3}X^{2}Y^{2}}\\ &\times \left[\beta_0\,\left(\,\frac{1}{2}\ln[X^{2}\mu^2]+\frac{1}{2}\ln[ Y^{2}\mu^2]\,\right)\right.\left.+\left(\frac{67}{9}\,-\,\frac{\pi^{2}}{3}\right)N_{c}-\frac{10}{9}N_{f}\right].
 \end{split}\end{equation}
 Here the UV cutoff $\mu^2$ is related to the  normalization point in the $\overline{MS}$ scheme $\mu_{\overline{MS}}^{2}$ (which we consider as being taken in the far UV) as $\mu^2=\frac{\mu_{\overline{MS}}^2}{4}e^{2\gamma}$ and $\beta_0$ is the first coefficient of the QCD $\beta$-function:
\begin{equation}\label{beta}
\beta_0\,=\beta_0^g+\beta_0^q\equiv\,\frac{11N_{c}\,-\,2N_{f}}{3}\,.
 \end{equation}



  Next:
  \begin{eqnarray}\label{JSSJ}
&&\hspace{-1cm}K_{JSSJ}(\mathbf{x},\, \mathbf{y},\, \mathbf{z},\, \mathbf{z}^{\prime})\nonumber\\
&&=\frac{\alpha_{s}^{2}(\mu)}{16\pi^{4}}\left[\frac{4}{Z^{4}}+\left\{ 2\frac{X^{2}(Y^{\prime})^{2}+(X^{\prime})^{2}Y^{2}-4(X-Y)^{2}Z^{2}}{Z^{4}(X^{2}(Y^{\prime})^{2}-(X^{\prime})^{2}Y^{2})}+\frac{(X-Y)^{4}}{X^{2}(Y^{\prime})^{2}-(X^{\prime})^{2}Y^{2}}\right.\right.\nonumber\\
&&\left.\times\left(\frac{1}{X^{2}(Y^{\prime})^{2}}+\frac{1}{Y^{2}(X^{\prime})^{2}}\right)+\frac{(X-Y)^{2}}{Z^{2}}\left(\frac{1}{X^{2}(Y^{\prime})^{2}}-\frac{1}{Y^{2}(X^{\prime})^{2}}\right)\right\} \ln\left(\frac{X^{2}(Y^{\prime})^{2}}{(X^{\prime})^{2}Y^{2}}\right)\nonumber\\
&&\left.-\frac{2I(\mathbf{x},\,\mathbf{z},\,\mathbf{z}^{\prime})}{Z^{2}}-\frac{2I(\mathbf{y},\,\mathbf{z},\,\mathbf{z}^{\prime})}{Z^{2}}\right]+\widetilde{K}(\mathbf{x},\,\mathbf{y},\,\mathbf{z},\,\mathbf{z}^{\prime}),\end{eqnarray}
with $Z_i\equiv \mathbf{z}_i-\mathbf{z}^\prime_i$ and
\begin{eqnarray}\label{ktild2}
&&\widetilde{K}(\mathbf{x},\, \mathbf{y},\, \mathbf{z},\, \mathbf{z}^{\prime})=\frac{\alpha_{s}^{2}(\mu)}{16\pi^{4}}\left(\frac{(Y^{\prime})^{2}}{(X^{\prime})^{2}Z^{2}Y^{2}}-\frac{Y^{2}}{Z^{2}X^{2}(Y^{\prime})^{2}}+\frac{1}{Z^{2}(Y^{\prime})^{2}}-\frac{1}{Z^{2}Y^{2}}+\frac{(X-Y)^{2}}{X^{2}Z^{2}Y^{2}}\right.\nonumber\\
&&\quad\quad \left.-\frac{(X-Y)^{2}}{(X^{\prime})^{2}Z^{2}(Y^{\prime})^{2}}+\frac{(X-Y)^{2}}{(X^{\prime})^{2}X^{2}(Y^{\prime})^{2}}-\frac{(X-Y)^{2}}{X^{2}(X^{\prime})^{2}Y^{2}}\right)\ln\left(\frac{X^{2}}{(X^{\prime})^{2}}\right)+\left(\mathbf{x}\leftrightarrow \mathbf{y}\right).\end{eqnarray}\\
The term $\tilde K$ arises from a specific ordering of the rotation operators $J_{R(L)}$ in the terms we  omitted in Eq.~\eqref{ham}. Similarly to those terms, $\tilde K$
does not contain any physics related to the QCD $\beta$-function and will not play any role in our discussion.

The terms involving $I(\mathbf{x})$ and $I(\mathbf{y})$ do not depend on one of the variables that is integrated over, either $\mathbf{x}$ or $\mathbf{y}$. These terms therefore vanish when the Hamiltonian acts on globally color singlet states. The explicit form for them  is given by 
\begin{equation}\begin{split}
&I(\mathbf{x},\,\mathbf{z},\,\mathbf{z}^{\prime})\,\equiv\,\frac{1}{X^{2}-(X^{\prime})^{2}}\left(\frac{X^{2}+(X^{\prime})^{2}}{Z^{2}}\,-\,\frac{X\cdot X^{\prime}}{X^{2}}\,-\,\frac{X\cdot X^{\prime}}{(X^{\prime})^{2}}\,-\,2\right)\ln\left(\frac{X^{2}}{(X^{\prime})^{2}}\right)\\
&\quad\quad =\,\frac{1}{X^{2}-(X^{\prime})^{2}}\left(\frac{X^{2}+(X^{\prime})^{2}}{Z^{2}}\,+\,\frac{Z^{2}-X^{2}}{2(X^{\prime})^{2}}\,+\,\frac{Z^{2}-(X^{\prime})^{2}}{2X^{2}}\,-\,3\right)\ln\left(\frac{X^{2}}{(X^{\prime})^{2}}\right).\\
\end{split}\end{equation}
For further reference we note that at  small values of $|X|\ll |Y|$, 
\begin{align}\label{intk}
\int_{\mathbf{z}^{\prime}}[K_{JSSJ}-\tilde K]
\approx 
\frac{\alpha_{s}^{2}}{4\pi^{3}} \frac{11}{3} \frac{X\cdot Y}{X^2 Y^2} \ln X^2 \mu^2 \,.
\end{align}
Finally 
\begin{eqnarray}\label{kqq}
K_{q\bar{q}}(\mathbf{x},\,\mathbf{y},\,\mathbf{z},\,\mathbf{z}^{\prime})&&=\frac{\alpha_{s}^{2}N_{f}}{8\pi^{4}}\left(\frac{2}{Z^{4}}-\frac{(X^{\prime})^{2}Y^{2}+(Y^{\prime})^{2}X^{2}-(X-Y)^{2}Z^{2}}{Z^{4}\left(X^{2}(Y^{\prime})^{2}-(X^{\prime})^{2}Y^{2}\right)}\ln\left(\frac{X^{2}(Y^{\prime})^{2}}{(X^{\prime})^{2}Y^{2}}\right)\right.\nonumber\\
&&\left.-\frac{I_{f}(\mathbf{x},\,\mathbf{z},\,\mathbf{z}^{\prime})}{Z^{2}}-\frac{I_{f}(\mathbf{y},\,\mathbf{z},\,\mathbf{z}^{\prime})}{Z^{2}}\right),\end{eqnarray}
where the term that vanishes when acting on color singlet states is:
\begin{equation}\begin{split}&I_{f}(\mathbf{x},\, \mathbf{z},\, \mathbf{z}^{\prime})\,\equiv\,\frac{2}{Z^{2}}-\frac{2X\cdot X^{\prime}}{Z^{2}(X^{2}-(X^{\prime})^{2})}\ln\left(\frac{X^{2}}{(X^{\prime})^{2}}\right).\\
\end{split}\end{equation}

\subsection{\label{sec:1} The ``subtraction'' method}
Importantly,  the term involving the kernel $K_{JSJ}$ has the very same structure as the LO Hamiltonian. It has an explicit  logarithmic UV divergence in the form of $\ln \mu_{\overline{MS}}$. The coefficient multiplying  $\ln \mu_{\overline{MS}}$ is twice the value necessary to collect these logarithms into the QCD $\beta$-function (see e.g. Ref.~\cite{Kovchegov:2012mbw}). 
However
the kernel $K_{JSSJ}(\mathbf{x},\,\mathbf{y},\,\mathbf{z},\,\mathbf{z}^{\prime})$ also  leads to logarithmic UV divergence when the distance between the two emitted gluons vanishes, $Z=\mathbf{z}-\mathbf{z'}\rightarrow 0$ as was noted in Refs.~\cite{Balitsky:2006wa,Kovchegov:2006vj}. The same is true for the $q\bar q$ kernel $K_{q\bar q}$. For simplicity of presentation we will discuss explicitly only the gluon emission for now. The $q\bar q$ emission is treated in an identical way. Since at short distances $\left.S^\dagger(\mathbf{z'})S(\mathbf{z}) \right| _{\mathbf{z'}\rightarrow \mathbf {z}}\rightarrow 1$, one can get rid of the divergence in the $K_{JSSJ}$ term by subtracting from it a linear term in $S$, and adding the same term to $K_{JSJ}$ term. This results in a UV finite  term involving the second power of $S$ at the expense of adding a piece to the $K_{JSJ}$ kernel.  
The coefficient of the $\ln \mu^2_{\overline{MS}}$ in the modified $K_{JSJ}$ is then exactly right to absorb this term into the $\beta$-function. 

However there is arbitrariness in the way one can subtract the divergence form $K_{JSSJ}$. 
For example in Ref.~\cite{Balitsky:2006wa}, the divergent subtraction term was multiplied by the Wilson line at position $S(\mathbf{z})$, while in Ref.~\cite{Kovchegov:2006vj} the position of the Wilson line was determined by the kinematics of DGLAP splitting in coordinate space. 
Since the subtracted term is then added to $K_{JSJ}$, the result is not universal. For example, in Ref.~\cite{Balitsky:2006wa}, the result\footnote{Note that in Ref.~\cite{Balitsky:2006wa} and Ref.~\cite{Kovchegov:2006vj}, 
the dipole kernel was considered, i.e. the action of JIMWLK on a dipole color singlet. 
The kernels are trivially related through (see e.g. Ref.~\cite{Kovner:2014lca})
\begin{align}
K_{\rm dipole} (X,Y)
= 
K_{JSJ} (X,X) + K_{JSJ} (Y,Y) - 2 K_{JSJ}(X,Y)     \,. 
\end{align}
} of the subtraction (we write out explicitly only the $\beta_0$-dependent terms)  is~\footnote{In terms of the kernels, the subtraction is given by (see Ref.~\cite{Lublinsky:2016meo}) 
\begin{equation}\label{jsjsub}
K_{JSJ} \to K^{\rm sub}_{JSJ}(\mathbf{x},\, \mathbf{y},\, \mathbf{z})\,\equiv\, K_{JSJ}(\mathbf{x},\, \mathbf{y},\, \mathbf{z})\,
-\frac{N_{c}}{2}\,\int_{\mathbf{z}^{\prime}}\, K_{JSSJ}(\mathbf{x},\, \mathbf{y},\, \mathbf{z},\, \mathbf{z}^{\prime})
-\frac{1}{2}\,\int_{\mathbf{z}^{\prime}}\, K_{q\bar q}(\mathbf{x},\, \mathbf{y},\, \mathbf{z},\, \mathbf{z}^{\prime}).
  \end{equation}}
\begin{equation}
K_{JSJ}\rightarrow K^B_{JSJ}=
\frac{\alpha_s^2(\mu) \beta_0}{ 16\pi^3}\left\{
-  \frac{(X-Y)^2}{X^2 Y^2}\ln(X-Y )^2\mu^2
+
\frac{1}{X^{2}}\ln Y^{2}\mu^{2}+\frac{1}{Y^{2}}\ln X^{2}\mu^{2}\right\} 
\end{equation}
while in Ref.~\cite{Kovchegov:2006vj}
\begin{equation}
K_{JSJ}\rightarrow K^{KW}_{JSJ}=
\frac{\alpha_s^2(\mu) \beta_0}{ 8\pi^3}
\frac{X\cdot Y} {X^2 Y^2}
\left\{
\frac{X^2 \ln X^2 \mu^2 - Y^2 \ln Y^2 \mu^2} {X^2-Y^2}
- 
\frac{X^2 Y^2}{X\cdot Y} \frac{\ln \frac{X^2}{Y^2} }{X^2-Y^2} 
\right\} \,.
\end{equation}
As the next step both in Refs.~\cite{Balitsky:2006wa} and \cite{Kovchegov:2006vj} one chooses to absorb {\it all logarithmic terms} in $K_{JSJ}$ into  renormalized coupling constant. The scale in $\alpha_s(\Lambda)$ is chosen in such a way as to reproduce all the logarithms in $K_{JSJ}$ {\it including their finite parts} when expanding $\alpha_s(\Lambda)$ to NLO in $\alpha_s(\mu)$. The two subtraction prescriptions therefore result in different resummations, corresponding to different finite NLO leftovers in $K_{JSSJ}$ (below by compact $\alpha_s(X^2)$ we obviously mean $\alpha_s(X^{-2})$):
\begin{eqnarray}
&&K^B_{JSJ}\rightarrow
\frac{\alpha_s((X-Y)^2)}{2\pi^2} \frac{X \cdot Y}{X^2 Y^2} \notag \\ &
   & \quad \quad \quad + 
    \frac{\alpha_s(X^2)}{4 \pi^2} \frac{1}{ X^2} \left(1 - \frac{\alpha_s((X-Y)^2)}{\alpha_s(Y^2)} \right)
    + 
    \frac{\alpha_s(Y^2)}{4 \pi^2} \frac{1}{ Y^2} \left(1 - \frac{\alpha_s((X-Y)^2)}{\alpha_s(X)} \right),
\\ &&\notag \\
&&K^{KW}_{JSJ}\rightarrow
    \frac{1}{2\pi^2}
    \frac{\alpha_s(X^2) \alpha_s(Y^2)} {\alpha_s(R^2)}\frac{X\cdot Y} {X^2 Y^2}, 
\end{eqnarray}
where, in the latter, a rather complex scale was introduced in the denominator 
\begin{align}
    R^2 = \sqrt{X^2 \,Y^2} \left( \frac{Y^2}{X^2}\right)^{\Theta/2}, \quad\quad\quad
    \Theta = 
    \frac{X^2  + Y^2} {X^2-Y^2} 
    - 2 \frac{X^2 Y^2}{X\cdot Y} \frac{1} {X^2-Y^2}\,.
\end{align}

The subtraction prescriptions described above are rather {\it ad hoc}. Although both render the kernels finite, one can devise many more subtraction prescriptions that achieve the same goal. Critically, there is no obvious reason for the resummation of the finite logarithmic parts of $K_{JSJ}$ that do not contain the UV cutoff $\mu_{\overline{MS}}$ into the running coupling in this or that particular way. 

In the next section, we give a transparent physical argument to the effect that the correct way to resum the transverse logarithms is rather different, and in addition to the QCD running coupling it requires the resummation of DGLAP splittings in the JIMWLK Hamiltonian.

\section{\label{dressing}The dressed gluon and the DGLAP splittings}
\label{Sec3}
We start by reexamining the origin of the transverse logarithm in Eq.~\eqref{kprime}.

As we alluded to before, in this term the coefficient of the UV divergent contribution  $\ln \mu^2_{\overline{MS}}$ is $2\beta_0$ (relative to the LO), and thus cannot be simply resummed into the running QCD coupling in the LO term. The reason for this is easy to understand. Recall, that in QCD one can define  the running coupling as  the matrix element of the interaction Hamiltonian between physical {\it dressed} gluon states. On the other hand, $K_{JSJ}$ is calculated as the amplitude for production of a {\it bare} gluon state from the valence charge \cite{Lublinsky:2016meo}. To relate this amplitude to the coupling constant one needs to multiply the gluon state by the wave function renormalization factor ${\cal Z}^{1/2}(Q^2)$, which at one loop is given by 
\begin{equation}\label{Z}
{\cal Z}^{1/2}(Q^2)=1+\frac{\alpha_s}{8\pi}\beta_0\ln\frac{Q^2}{\mu^2}\,.
\end{equation}
The scale in Eq.~\eqref{Z} is an arbitrary normalization scale at which one chooses to define the renormalized gluon field 
\begin{equation}\label{rf}
A_\mu^Q(x)={\cal Z}^{-1/2}(Q)A_{\mu}(x)\,.
\end{equation}
Thus to identify the UV logarithm associated with the running of the coupling one has to express the bare gluon field (or state) in terms of the ``renormalized'' gluon field via Eq.~\eqref{rf}.
Technically within the JIMWLK Hamiltonian this procedure amounts to  multiplying the LO kernel by the factor ${\cal Z}^{1/2}(Q)$, with the understanding that now the production and scattering refer to the dressed renormalized gluon. This multiplication of the LO kernel absorbs some of the NLO corrections and hence results in the modification of the NLO kernel\footnote{In this discussion we only consider the coefficient of the real term (the $JSJ$ term) for simplicity.
}
\begin{equation}\begin{split}\label{kmod}
 K_{JSJ}\rightarrow  &\frac{\alpha_{s}^{2}X\cdot Y}{4\pi^{3}X^{2}Y^{2}}\,\left(\beta_0\,\left[\,\frac{1}{2}\ln[X^{2}\mu^{2}]\,+\,\frac{1}{2}\ln[Y^{2}\mu^2]\,-\frac{1}{2}\ln\frac{\mu^2}{Q^2}\right]\right.\left.+\left(\frac{67}{9}-\frac{\pi^{2}}{3}\right)N_{c}\,-\,\frac{10}{9}N_{f}\right).
 \end{split}\end{equation}
This expression now clearly contains all the UV logs to be combined into the running coupling. 

However now the question arises, what is the fate of the UV divergence in $K_{JSSJ}$ discussed above?
The answer is that this divergence should also cancel if one rewrites the JIMWLK Hamiltonian in terms of the physical dressed gluon amplitudes. The reason this has not been apparent so far, is because in our discussion above we have only included the simple multiplicative wave function renormalization factor into the ``redefinition'' of the dressed gluon. However at NLO the physical dressed gluon state also contains a two gluon (and a quark-anti quark) component due to gluon splitting. We have to include those in the definition of the dressed gluon scattering amplitude. 
This is the goal of the rest of this section.

Technically we take the following route. At NLO a gluon can split into a two-gluon (or quark-antiquark) pair. This splitting in collinear setting is described by the DGLAP splitting function. The splitting amplitude is of course well known. The square of this amplitude provides the real contribution to the $g\rightarrow gg \ (\bar q q)$  DGLAP splitting functions. The  virtual part of the splitting function is given by the total probability that the initial bare gluon state decays into a $gg$ (or $\bar q q$) state. 

Our goal is to write the scattering amplitude of the dressed state which is resolved with transverse resolution $Q$. In terms of Wilson renormalization group that would correspond to ``integrating out'' modes with transverse momenta above the resolution scale $Q$ which defines the ``renormalized'' dressed states, as in the discussion above. We assume, as always that individual bare partons scatter on the target eikonally. 

\subsection{The dressed gluon scattering matrix.}

We consider first the splitting into two gluons, and in this section disregard the possibility of splitting into $\bar q q$ for simplicity. It is straightforward to include the quarks, and we will do it in Section \ref{Sec5}.

The $S$-matrix of the gluon state dressed at order $\alpha_s$ with resolution $Q$ then equals 
\begin{align}\label{sfat}
    \mathbb{S}_Q^{ab}(\mathbf{z}) \,= 
    S^{ab}(\mathbf{z}) & +\, \frac{\alpha_s} {2\pi^2} \int_0^1\frac{d\xi}{\xi_+(1-\xi)_+} \left( \xi^2+ (1-\xi)^2 + \xi^2(1-\xi)^2\right) \nonumber  \\& \times \int \limits_{\mu^{-1}<|Z|<Q^{-1}}\frac{d^2Z}{Z^2}
        \left[ D^{ab}(\mathbf{z} + (1-\xi) Z, \mathbf{z} -\xi Z) - N_c S^{ab}(\mathbf{z})  \right]
\end{align}
This definition follows from the kinematics of gluon splitting in coordinate space, where the distance between the daughter gluons and the parent gluon is determined by the longitudinal momentum fraction $\xi$ carried by each gluon. The first term ($D$)
in the second line in (\ref{sfat}) represents real emission, while the second term is the virtual correction.
Note that when coordinates in $D$ equal each other it simplifies to 
\begin{align}
      D^{ab}(\mathbf{z}, \mathbf{z}) = 
    {\rm Tr}[ \underbrace{S^+(\mathbf{z}) T^a  S(\mathbf{z})}_{S_{a a'} T^{a'}}  T^b ] = N_c S^{ab} (\mathbf{z})\,. 
\end{align}

Importantly the plus prescription on the integral is defined with respect to both poles, at $\xi=0$ and $\xi=1$, that is 
\begin{align}\label{xi++}
    \int_0^1 \frac{d\xi}{\xi_+(1-\xi)_+} 
    f(\xi)  = 
    \int_0^1 d\xi  
    \left( 
    \frac{1}{\xi(1-\xi)} f(\xi)
    -
    \frac{1}{\xi} 
    f(0)
    - 
    \frac{1}{1-\xi} 
    f(1)
    \right)\,.
\end{align}
As we will see later, this double subtraction is needed to correctly account for the large transverse logarithm in $K_{JSSJ}$.

This is in contradistinction to the usual DGLAP splitting function, where only the $\xi=0$ pole is treated with the plus prescription. The reason for this double subtraction of both poles from the splitting function is that both  poles are already included in the energy evolution generated by the LO JIMWLK equation (in \cite{Lublinsky:2016meo} these poles are referred to as LO$^2$ terms). 
It is only the finite remainder of the splitting function that has to be included in ``collinear splitting term'' we currently consider. In this sense the use of the term ``DGLAP" is a slight abuse of language. We will refer to it therefore as ``DGLAP-like" most of the time, although sometimes our terminology will slip.  It is important however to keep in mind this point, and we will come back to it in the Section \ref{Sec6}.

With the definition Eq.~\eqref{xi++} we have
\begin{align}\label{integral}
    \int_0^1 \frac{d\xi}{\xi_+(1-\xi)_+} \left( \xi^2+ (1-\xi)^2 + \xi^2(1-\xi)^2\right) = -\frac{11}{6}=-\frac{\beta^g_0}{2N_c}
\end{align}
which is proportional to the one loop gluon contribution to the $\beta$- function, $\beta^g_0=\frac{11N_c}{3}$. 

The next step is now clear. We should perturbatively re-express all the eikonal factors in the Hamiltonian in terms of $\mathbb{S}_Q$, using 
\begin{align}\label{ssq}
  S^{ab}(\mathbf{z})  \, = \,\mathbb{S}_Q^{ab}(\mathbf{z}) 
   &-\,\frac{\alpha_s} {2\pi^2} \int_0^1\frac{d\xi}{\xi_+(1-\xi)_+} \left( \xi^2+ (1-\xi)^2 + \xi^2(1-\xi)^2\right) \nonumber  \\&  \times \int \limits_{|Z|<Q^{-1}}\frac{d^2Z}{Z^2}
        \left( \mathbb{D}_Q^{ab}(\mathbf{z} + (1-\xi) Z, \mathbf{z} -\xi Z) - N_c \mathbb{S}_Q^{ab}(\mathbf{z})  \right)\,.
\end{align}
 with
 \begin{align}
    \mathbb{D}_Q^{ab}(\mathbf{z}_1, \mathbf{z}_2) = 
    {\rm Tr}[ T^a  \mathbb{S}_Q(\mathbf{z}_1)  T^b \mathbb{S}_Q^+(\mathbf{z}_2)]\,.
\end{align}
We note that
\begin{equation}
        \frac{\alpha_sN_c} {2\pi^2} \int_0^1 \frac{d\xi}{\xi_+(1-\xi)_+} \left( \xi^2+ (1-\xi)^2 + \xi^2(1-\xi)^2\right) \int \limits_{|Z|<Q^{-1}}\frac{d^2Z}{Z^2} \mathbb{S}_{ab}(\mathbf{z})=-\frac{\alpha_s\beta_0^g}{4\pi}\ln\frac{\mu^2}{Q^2}\mathbb{S}^{ab}(\mathbf{z})\,.
        \end{equation}
        
At this point, we will simplify our derivations in the following by neglecting the $\xi$ dependence in the coordinates of the Wilson lines in Eq.~\eqref{sfat}. Recall  that $\xi=0$ and $\xi=1$ do not play any special role in the integration over $\xi$ due to the double plus prescription. It is clear that with the logarithmic accuracy the exact location of the two points $\mathbf {z}_1$ and $\mathbf{z}_2$ within the integration region in Eq.~\eqref{sfat} does not matter as long as the distance between them remains fixed. Since in this paper we only concern ourselves with leading logarithmic contributions, we will simply set $\xi$ to a fixed value in the arguments of the Wilson lines. To preserve the symmetry between $\xi$ and $1-\xi$ we substitute $\xi=1/2$ in the coordinates of $D$ in Eq.~\eqref{sfat} and onward. This allows us to perform explicitly the integral over $\xi$ using Eq.~\eqref{integral}.

Thus from now on we use
\begin{align}\label{sfat1}
    \mathbb{S}_Q^{ab}(\mathbf{z}) &= 
     [1+\frac{\alpha_s\beta_0^g}{4\pi}\ln\frac{\mu^2}{Q^2}]S^{ab}(\mathbf{z}) - \frac{\alpha_s\beta_0^g} {4\pi^2N_c} \int \limits_{|Z|<Q^{-1}}\frac{d^2Z}{Z^2}
         D^{ab}(\mathbf{z} +  Z/2, \mathbf{z} -Z/2)   
\end{align}
and 
\begin{align}\label{sfat2}
    S^{ab}(\mathbf{z}) &= 
     [1-\frac{\alpha_s\beta_0^g}{4\pi}\ln\frac{\mu^2}{Q^2}]\mathbb{S}_Q^{ab}(\mathbf{z}) +\frac{\alpha_s\beta_0^g} {4\pi^2N_c} \mathbb{D}^{ab}_Q(\mathbf {z})\notag \,.
\end{align}
 where we have introduced the notation
\begin{equation}
 \mathbb{D}^{ab}_Q(\mathbf {z})\equiv  \int \limits_{|Z|<Q^{-1}}\frac{d^2Z}{Z^2} \mathbb{D}^{ab}(\mathbf{z} +  Z/2, \mathbf{z} - Z/2)\,.
   \end{equation}      
Again, we stress that this simplification can be relaxed without 
altering logarithmic in $Q$ terms.  

Since at the moment, we  work at fixed order $\alpha_s^2$, the substitution of $\mathbb{S}_Q$ should be done perturbatively, which in practice means that the $O(\alpha_s)$ term in Eq.~\eqref{ssq}  should only be kept at $H^{\rm LO}_{\rm JIMWLK}$, while in $H^{\rm NLO}_{\rm JIMWLK}$ all factors of $S$ should be substituted by $\mathbb{S}_Q$.

Let us first consider the real term in $H^{\rm LO}_{\rm JIMWLK}$. Substituting Eq.~\eqref{ssq} we obtain
\begin{eqnarray}\label{subst}
\frac{\alpha_{s}}{2\pi^{2}}\int_{\mathbf{x},\, \mathbf{y},\, \mathbf{z}}\frac{X\cdot Y}{X^{2}Y^{2}}\,\,2J_{L}^{a}(\mathbf{x})\, S^{ab}(\mathbf{z})\, J_{R}^{b}(\mathbf{y})&=&\\
&&\hspace{-2cm}=\frac{\alpha_{s}}{2\pi^{2}}\int_{\mathbf{x},\, \mathbf{y},\, \mathbf{z}}\frac{X\cdot Y}{X^{2}Y^{2}}\,\,2J_{L}^{a}(\mathbf{x})\, \mathbb{S}_{Q}^{ab}(\mathbf{z})\, J_{R}^{b}(\mathbf{y})\left[1-\frac{\alpha_s\beta^g_0}{4\pi}\ln\frac{\mu^2}{Q^2}\right]
\nonumber\\
&&\hspace{-2cm}+\frac{\alpha^2_{s}\beta_0^g}{8\pi^{4}N_c}\int_{\mathbf{x},\, \mathbf{y},\, \mathbf{z}}\frac{X\cdot Y}{X^{2}Y^{2}}2J_{L}^{a}(\mathbf{x})\, J_{R}^{b}(\mathbf{y})\mathbb{D}^{ab}_Q(\mathbf {z})\nonumber\,.
\end{eqnarray}
We observe that the logarithmic term linear in $\mathbb{S}$ in Eq.~\eqref{subst} combines with the real part of the kernel $K_{JSJ}$ to yield precisely the transverse logarithm in Eq.~\eqref{kmod}. In addition the quadratic in $\mathbb{S}$ term in Eq.~\eqref{subst} combines with the real term of $K_{JSSJ}$ in Eq.~\eqref{JSSJ}, so that the UV divergence at $\mathbf{z}-\mathbf{z}'\rightarrow 0$ cancels. This is precisely what we set out to achieve by introducing the dressed gluon state. 

It is thus very convenient to re-express the JIMWLK Hamiltonian in terms of the dressed gluon operators, as it achieves the same purpose as various subtractions discussed above, but with clear physical justification.

There is one subtlety in the procedure just described. What we did is to introduce a gluon state ``dressed'' by the DGLAP emissions with resolution $Q^2$ irrespective of whether the gluon in question is far away or close to the valence charge that it was emitted from. Physically this is not right. Proximity to the emitter provides an infrared cutoff on possible DGLAP splittings. The collinear logarithm in the standard DGLAP cascade is formally collinearly divergent for large distance or  small transverse momentum   emissions.  This collinear divergence however is there only as long as one assumes that the parton splits completely independently of all other partons present in the hadronic wave function. The moment one accounts for the fact that the gluon is not isolated in space, but is a part of hadronic wave function with finite size $R$, the collinear divergence disappears. This size $R$ can be a small perturbative scale for a heavy quarkonium, or a nonperturbative scale of order $\Lambda^{-1}_{\rm QCD}$ for a nucleon {\it etc}. Either way the presence of this scale regulates the DGLAP emissions in the IR by cutting off the phase space for splitting into pairs of the size larger than $R$.

 The situation is very similar in the case at hand. To see this we should  examine $K_{JSSJ}$ which reflects the probability of splittings present in JIMWLK at NLO. The probability of emission from a single source at point $\mathbf{x}$ can be read off the integrand of $K_{JSSJ}$ by setting $\mathbf{y}=\mathbf{x}$. 
 Setting $\mathbf{y}=\mathbf{x}$ and examining $K_{JSSJ}$ we find that for $Z^2\ll X^2$, the kernel behaves as $1/Z^2$. Integration over $Z$ in this range produces the UV divergence we discussed. On the other hand, when the size of the pair is large relative to the distance between the emitter and the center of mass of the pair, $Z^2\gg (X+\frac{Z}{2})^2$ we find that the kernel (probability of splitting) behaves as $(X+\frac{Z}{2})^2/Z^6$. One can see this also from eq.\eqref{intk} which makes it explicit that the integration over the size of the gluon pair is cut off by the distance to the closest source in the emission amplitude. This tells us that
$K_{JSSJ}$ does not allow DGLAP splittings of a gluon into pairs of transverse size larger than the distance between the gluon in question and its ``valence'' emitter. 

  In this discussion, the hadronic size $R$ is directly analogous to the distance $|X-Z/2|$ between the gluon and the closest parton in the wave function, most of the time, this closest parton  being the emitter of the gluon in question. 
  
 The procedure outlined in Eq.~\eqref{subst} does not differentiate between the two regimes $Q^{-2}<(X-Z/2)^2$ and $Q^{-2}>(X-Z/2)^2$ and therefore does not properly take into account the suppression discussed above. Mathematically the consequence is that, although when the last term in Eq.~\eqref{subst} is combined with the $K_{JSSJ}$ term in NLO JIMWLK it does get rid of the $Z\rightarrow 0$ divergence, it produces a logarithmic term $\ln(X^2Q^2)$, for all values of $X$, rather than only for $X^2>Q^{-2}$. This spurious logarithm at small $X$ may be large and its introduction should be avoided.
  
  It is not hard to modify our strategy sightly in order to avoid this difficulty. Let us define the resolution scale that depends on the position of the gluon relative to the closest valence charge. In the context of the real term in the LO JIMWLK, in terms of the coordinates 
  $\mathbf{x},\mathbf{y},\mathbf{z}$ we introduce:
  \begin{equation}
      \bar Q^2=\max\left\{Q^2, \frac{1}{X^2},\frac{1}{Y^2}\right\}\,.
  \end{equation}
 We now express $S(\mathbf{z})$ in terms of $\mathbb{S}_{\bar Q}(\mathbf{z})$ in $H^{\rm LO}_{\rm JIMWLK}$.

\subsection{The reorganized Hamiltonian}
The resulting Hamiltonian can be organized in the convenient way into two parts $H_Q^{J\mathbb{S}J}$ and $H_Q^{J\mathbb{SS}J}$.
To do so we first introduce an auxiliary scale 
\begin{equation}\label{tildeq}
   \tilde Q^2=\max\left\{\frac{1}{X^2},\frac{1}{Y^2}\right\}\,.
\end{equation}
We now split the virtual term in $H_{\rm JIMWLK}^{\rm LO}$  as follows
\begin{eqnarray}\label{split}
&&\int_{\mathbf{x,y,z}}\frac{X\cdot Y}{X^{2}Y^{2}}\left[ J_{L}^{a}(\mathbf{x}) J_{L}^{a}(\mathbf{y})+ J_{R}^{a}(\mathbf{x}) J_{R}^{a}(\mathbf{y})\right]=\int_{\mathbf{x,y,z}} \frac{X\cdot Y}{X^{2}Y^{2}}\notag \\
&& \times
\Bigg\{\left[1-\frac{\alpha_s\beta^g_0}{8\pi}(\ln X^2\mu^2+\ln Y^2\mu^2)+ \frac{\alpha_s\beta^g_0}{8\pi}(\ln X^2\tilde Q^2+\ln Y^2\tilde Q^2)\right] \left[J_{L}^{a}(\mathbf{x}) J_{L}^{a}(\mathbf{y})\,+ J_{R}^{a}(\mathbf{x}) J_{R}^{a}(\mathbf{y})\right]\nonumber  \\
&&+\,\frac{\alpha_s\beta^g_0}{4\pi}\ln \frac{\mu^2}{\tilde Q^2}\left[J_{L}^{a}(\mathbf{x})\, J_{L}^{a}(\mathbf{y})\,+\, J_{R}^{a}(\mathbf{x})\, J_{R}^{a}(\mathbf{y})\right]\Bigg\} \,.
\end{eqnarray} 
The logic of this splitting will become clear in a short while\footnote{One may wonder if one needs also to express the right and left rotation operator in terms of rotations associated with the dressed $S$-matrices $\mathbb{S}_Q$. It turns out, however, that even if done, this procedure is purely perturbative and does not contribute to the resummation of large logarithms, see Appendix A. It is therefore optional, and in a sense can be regarded as simply ``fixing the factorization scheme'' for DGLAP-like resummation. For simplicity, we choose not to do it here.}.

It is convenient to reorganize the Hamiltonian as 
\begin{equation}
H_{\rm JIMWLK}=H^{J\mathbb{S}J}_{Q}+H^{J\mathbb{SS}J}_{Q}+\ldots 
\end{equation}
where
\begin{eqnarray}\label{reorg1}
&&H^{J\mathbb{S}J}_{ Q}
=\int_{\mathbf{x,y,z}}\frac{\alpha_{s}}{2\pi^{2}}\,\frac{X\cdot Y}{X^{2}Y^{2}}\notag\\
&&\times\Bigg\{\left(1+\frac{\alpha_s\beta^g_0}{8\pi}\left[\ln[X^{2}\mu^{2}]+\,\ln[Y^{2}\mu^2]\right]\right)\left[J_{L}^{a}(\mathbf{x}) J_{L}^{a}(\mathbf{y})+ J_{R}^{a}(\mathbf{x})\, J_{R}^{a}(\mathbf{y})-2J_{L}^{a}(\mathbf{x}) \mathbb{S}_{\bar Q}^{ab}(\mathbf{z}) J_{R}^{b}(\mathbf{y})\right]\nonumber\\
&&\,+\,\frac{\alpha_s\beta^g_0}{8\pi}\Big[(\ln X^2\tilde Q^2+\ln Y^2\tilde Q^2)\, \left[J_{L}^{a}(\mathbf{x})\, J_{L}^{a}(\mathbf{y})\,+\, J_{R}^{a}(\mathbf{x})\, J_{R}^{a}(\mathbf{y})\right]\nonumber\\
&&\hspace{7cm}-2
(\ln X^2\bar Q^2+\ln Y^2\bar Q^2)J_{L}^{a}(\mathbf{x})\, \mathbb{S}_{\bar Q}^{ab}(\mathbf{z})\, J_{R}^{b}(\mathbf{y})\,\Big]\Bigg\}\,.
\end{eqnarray}
\begin{eqnarray}\label{reorg2}
 &&H^{J\mathbb{SS}J}_Q =\int_{\mathbf{x},\, \mathbf{y},\mathbf{z}, \mathbf{z}^{\prime}}\,\hspace{-0.3cm} K_{JSSJ}(\mathbf{x},\mathbf{y},\mathbf{z}, \mathbf{z}^{\prime})J_{L}^{a}(\mathbf{x})\mathbb{D}_{\bar Q}^{ad}(\mathbf{z},\mathbf{z}^{\prime}) J_{R}^{d}(\mathbf{y})
 -\frac{\alpha^2_{s}\beta^g_0}{4\pi^{3}}\int_{\mathbf{x},\, \mathbf{y},\mathbf{z}}\frac {X\cdot Y}{X^{2}Y^{2}}
 \mathbb{D}^{ab}_{\bar Q}(\mathbf {z})J_L^a(\mathbf{x})J_R^b(\mathbf{y})\nonumber\\
 &&
 -\left[\frac{N_c}{2}\int_{\mathbf{x},\, \mathbf{y},\mathbf{z}, \mathbf{z}^{\prime}}\hspace{-0.3cm} K_{JSSJ}(\mathbf{x},\mathbf{y},\mathbf{z}, \mathbf{z}^{\prime})
 -\frac{\alpha^2_{s}\beta^g_0}{8\pi^{3}}\int_{\mathbf{x},\, \mathbf{y},\mathbf{z}} \frac{X\cdot Y}{X^{2}Y^{2}}\,\ln\frac{\mu^2}{\tilde Q^2}\right]\left[J_R^a(\mathbf{x})J_R^a(\mathbf{y})+J_L^a(\mathbf{x})J_L^a(\mathbf{y})\right]\,. 
\end{eqnarray}
Note that Eq.~\eqref{reorg1} exhibits explicitly real-virtual cancellations. As $X\rightarrow 0$, we have $\tilde Q=\bar Q=1/|X|$, $\,\,\mathbb{S}_{\bar Q}(\mathbf{z})=S(\mathbf{z})\simeq S(\mathbf{x})$,
and therefore the real and virtual terms cancel for points $\mathbf{z}$ such that $|X|<Q_T^{-1}$.

Eq.~\eqref{reorg2} has some appealing properties. First, it is UV finite. The integral over $Z$ in the definition of $\mathbb{D}_{\bar Q}(\mathbf{z})$ cancels the UV divergence in the first term. The UV cancellation also occurs between the two virtual terms in the second line of Eq.~\eqref{reorg2}. 

Moreover, Eq.~\eqref{reorg2} contains no large finite logarithms if the target is dense, i.e. if $Q_T$ is in fact a saturation momentum.

For the real term we argue as follows. The JIMWLK approach is valid for scattering of a dilute projectile on a dense target. Let us assume that the target field distribution is characterized by a saturation momentum $Q_T$. Two eikonal scattering factors within a transverse distance $|Z|\ll (Q_{T})^{ -1}$ are practically the same, while if $|Z|\gg (Q_{T})^{-1}$, the two eikonal factors fluctuate independently and therefore the product averages to zero. As we will explain shortly, our interest will be for $Q<Q_{T}$. Under these assumptions, the integral over the region $|Z|<(Q_{T})^{-1}$ cancels between the first two terms in the first line by the same token as does the UV cutoff $\mu$. For the distances $Q^{-1}>|Z|> (Q_{T})^{-1}$ there is no algebraic cancellation, however each one of the terms is small since it contains two eikonal factors separated by a large distance, and this product averages to zero. This range of $Z$ therefore does not give a logarithmic contribution to the hadronic $S$-matrix. Thus the contribution only comes from a small region in the phase space $|Z|\sim (Q_{T})^{-1}$ and cannot be logarithmically large for any $Q<Q_T$.

As for the virtual term, integrating $K_{JSSJ}$ over $Z$, for $X^2<Y^2$ one obtains   $\ln\mu^2 X^2$, since the integration is cut off in the IR by $|X|$\footnote{The cutoff is actually $|Y|$ if $|Y|<|X|$, but these are mapped into each other by a simple change of variables $\mathbf{x}\rightarrow\mathbf{y}$.}. The coefficient of this logarithm is exactly the same as in the second virtual term, and the two logarithms cancel between the two virtual terms. 

One could worry that Eq.~\eqref{reorg1} contains $\ln\frac{Y^2}{X^2}$, which for $Y^2\gg X^2$ is  large and may require special attention. However this potential logarithm is harmless. It arises in the term responsible for emission of a virtual gluon very close to one of the valence charges in the wave function. The probability of emission of a gluon at any point $\mathbf{z}$ is given by the square of the sum of the emission amplitudes from all the valence charge. If $\mathbf{z}$ is much closer to one valence charge than to any of the others, the emission amplitude is dominated by the amplitude of emission from this particular valence charge, and the contribution of the other charges can be neglected. The same of course is also true for the conjugate amplitude. Thus the conjugate amplitude of such (virtual or real) emission is dominated by the emission from the same charge as in the amplitude. In other words, if $\mathbf{z}$ is very close to a given charge $\mathbf{x}$, the integral over $Y$ in the conjugate amplitude will be dominated by $Y\approx X$ simply due to the structure of the WW field in the kernel. Thus one does not expect that the region of $Y^2\gg X^2$ will be of any practical importance and $\ln\left(\frac{Y^2}{X^2}\right)$ should not be considered large.


So far we have just rewritten the NLO JIMWLK Hamiltonian in a convenient way. As argued above, the term $H^{J\mathbb{SS}J}_Q$ is UV finite and does not contain any large logarithms. All the large logarithms are now explicit in $H^{J\mathbb{S}J}_Q$. These logarithms come in two types. The first one is the UV divergent term in the first term of $H^{J\mathbb{SS}J}_Q$, and the second one is the finite log of the type $\ln (Y^2Q^2)$ in the last term. 

\section{The resummation}
\label{Sec4}
The two large logarithms, $\ln (X^2\mu^2)$  and $\ln (X^2Q^2)$
need to be resummed.
The physical origin of the two is quite clear: the UV logarithm is  associated with setting the scale of the running coupling, while the $Q^2$ dependent logarithm is associated with setting the resolution scale for DGLAP splittings.

\subsection{The running coupling}

The resummation of the UV logarithm is equivalent to choosing the scale in the running coupling. The simplest and the most natural scale choice for a Coulomb field at the distance $X$ from the source that creates it, is of course just $X$ itself. That is how things work in the textbook examples in QED. Indeed, Eq.~\eqref{reorg1} suggests precisely this scale choice. Defining
as usual, the running QCD coupling constant $\alpha_s(\mu^2)$ as the solution of the one loop renormalization group equation, we can rewrite Eq.~\eqref{reorg1}\footnote{The combination of the running couplings that reproduces the first order terms can be written as  
\begin{align}
\frac{g(X^{-2})g(Y^{-2})}{8\pi^3} \to \frac{\alpha^\lambda_s(X^{-2})\,\alpha^\lambda_s(Y^{-2})  \alpha^{1-2 \lambda}_s((X Y)^{-1}) } {2\pi^2}
\end{align}
with $\lambda$ - any real number. It is not uniquely defined at the NLO order.  The choice  $\lambda=1$ leads to the ``triumvirate'' form, see Refs.~\cite{Braun:1994mw, Levin:1994di,Chirilli:2013kca}, while our choice corresponds to $\lambda=1/2$. We find this last option  most natural.} in terms of the effective charge $g(\mu^2)$ defined as
$\alpha_s(\mu^2)={g^2(\mu^2)/4\pi}$,

\begin{eqnarray}\label{alpharesummed}
H^{J\mathbb{S}J}_{ Q}
&=&\int_{\mathbf{x,y,z}}\frac{g(X^{-2})g(Y^{-2})}{8\pi^3}\frac{X\cdot Y}{X^{2}Y^{2}}\,\\
&&\times\Bigg\{\, \left(1+ \frac{\alpha_s\beta^g_0}{8\pi}\left[\ln X^2\tilde Q^2+\ln Y^2\tilde Q^2\right]\right) \left[J_{L}^{a}(\mathbf{x})\, J_{L}^{a}(\mathbf{y})\,+\, J_{R}^{a}(\mathbf{x})\, J_{R}^{a}(\mathbf{y})\right]\,\nonumber\\
&&-\,2\left(1+ \frac{\alpha_s\beta^g_0}{8\pi}\left[\ln X^2\bar Q^2+\ln Y^2\bar Q^2\right]\right) J_{L}^{a}(\mathbf{x})\, \mathbb{S}_{\bar Q}^{ab}(\mathbf{z})\, J_{R}^{b}(\mathbf{y})\,\Bigg\}\nonumber\,.
\end{eqnarray}


\subsection{The DGLAP like corrections}
We rewrote the Hamiltonian in terms of $\mathbb{S}_Q$ in the fixed order in perturbation theory. This 
eliminated 
the original UV divergences, but introduced dependence on an arbitrary scale $Q$. This scale clearly plays the same role as the factorization scale in usual DIS. The physics (the Hamiltonian) eventually does not depend on this scale, but choosing a convenient value for it has the potential of simplifying calculations considerably. 

Varying the scale $Q$ does not change physics, but shifts large perturbative corrections between the $Q$ dependence of the Hamiltonian and the $Q$ dependence of the $S$-matrix of the dressed gluon. In particular, if we can choose the scale $Q$ such that the logarithm in Eq.~\eqref{alpharesummed} is not large, the Hamiltonian has a simple form that only perturbatively differs from the LO Hamiltonian (modulo the running coupling effects). The price we pay is that at this scale the dressed gluon is very different from the bare gluon state, as its $S$-matrix, $\mathbb{S}_Q(\mathbf{z})$ is very different from the original eikonal factor $S(\mathbf{z})$. 

This is closely analogous to DIS or jet production, where changing the resolution scale shifts large corrections between the PDF and the hard part of the scattering. In fact, the Hamiltonian $H_{\rm JIMLWK}$ is very similar to TMD PDF: in the leading order, the square of the WW field is precisely the number of soft gluons at point $\mathbf{z}$ emitted from the valence charges. The scattering matrix, on the other hand, is similar to the hard part. With high resolution $\mathbb{S}_{Q=\mu}=S$, which is analogous to the situation in DIS where choosing the factorization scale close to UV cutoff eliminates large corrections from the hard scattering part.  While for lower resolution scale $Q\ll\mu$, the dressed gluon scattering matrix, similarly to the hard scattering part, acquires large corrections.

In DIS and similar processes, the optimal choice of factorization scale is determined by the external scale furnished by the observable, e.g. the momentum transfer $Q^2$. In the case of  JIMWLK evolution, the role of this scale is played by the saturation momentum of the target, $Q_T$. Since the eikonal scattering factors $S(\mathbf{z}_1)$ and $S(\mathbf{z_2})$ do not differ from each other as long as $|\mathbf{z}_1-\mathbf{z}_2|<Q_T^{-1}$, for any choice of factorization scale $Q^2\ge Q_T^2$ the scattering matrix of a dressed gluon (which in coordinate space has a transverse size $\sim Q^{-1}$) is the same as the bare eikonal factor $S$ up to small perturbative corrections. The most convenient choice of $Q$ is, therefore, $Q^2=Q_T^2$, since it eliminates the necessity to evolve the dressed gluon density matrix. The Hamiltonian of course, has to be evolved to this scale, in a complete analogy to the DGLAP evolution of parton PDF's. Calculating this evolution is our next goal. 

Before doing so, however, we must generalize the evolution of the dressed gluon scattering matrix to account for possible further gluon emissions, which are essential if the evolution interval is logarithmically large.

Including additional DGLAP emissions in the dressed gluon scattering matrix is quite straightforward. We go back to Eq.~\eqref{sfat1} and write it in the differential form. We also use the fact that for an infinitesimal variation of $Q$, the change in the scattering matrix on the right-hand side of the equation can be neglected. We thus arrive at a DGLAP-like equation
\begin{equation}\label{dglaplike}
   \frac{\partial}{\partial \ln Q^2} \mathbb{S}_Q^{ab}(\mathbf{z}) = -
    \frac{\alpha_s\beta^g_0}{4\pi} 
    \left[ \mathbb{S}_Q^{ab}(\mathbf{z})-\frac{1}{N_c}\int \frac{d\phi}{2\pi}
        \left( \mathbb{D}_Q^{ab}(\mathbf{z} + \frac{1}{2} Q^{-1} \mathbf{e}_\phi, \mathbf{z} -\frac{1}{2} Q^{-1} \mathbf{e}_\phi)   \right) \right]
\end{equation}
where $\mathbf{e}_\phi$ is the unit vector in the tangential direction $\phi$.
The solution of this equation, in principle, sums all DGLAP emissions of the initial gluon within the transverse radius $Q^{-1}$.

The resummation of the DGLAP logarithms is achieved by requiring that the resummed Hamiltonian does not depend on the arbitrary scale $Q$ that we have introduced above:
\begin{equation}
    \frac{d}{d\ln Q^2}H_{\rm JIMLWK}=\frac{\partial H}{\partial \ln Q^2} 
+  \int_u \left[\frac{\delta H }{\delta \mathbb{S}_Q(u)}  \frac{\partial \mathbb{S}_Q(u)}{\partial \ln Q^2}
 \right]
=0\,. 
\end{equation}

To resum the DGLAP logarithms we need to solve the RG equation for the Hamiltonian. We will do this in two simple case below. 
But first, we have to understand the initial condition and the interval of the evolution in $Q$ over which we need to evolve.

As discussed above, we assume that the target is characterized by the correlation length $Q_T^{-1}$~\footnote{Solution of the LO JIMWLK shows that in the saturated regime,  the color correlation length is about 2.5 times larger than the inverse saturation momentum~\cite{Duan}.}. Clearly, if $Q>Q_T$, the DGLAP cascade we resume in $\mathbb{S}_Q$ is localized inside the single correlation length, the scattering amplitude of all gluons (and quarks) inside the cascade are the same, the cascade is not resolved by scattering on the target, and its scattering amplitude is equal to the single eikonal factor $S$. Thus it does not make sense to consider $Q$ greater than $Q_T$. Therefore the evolution in $Q$ starts from $Q=Q_T$ with the  initial condition 
\begin{equation}\label{init}
\mathbb{S}_{Q_T}(\mathbf{x})=S(\mathbf{x})\,.
\end{equation}
As for the interval of the evolution, clearly we should stop the evolution at the value of $Q$ for which the Hamiltonian, when expressed in terms of $\mathbb{S}_Q$ does not contain large logarithms. At this point $H_{\rm JIMWLK}$ will be a nice function of $\mathbb{S}_Q$ with coefficients in the expansion in powers of $\alpha_s$ being all numbers of order unity (not enhanced by transverse logs). 

Let us examine the expressions Eq.~\eqref{reorg1},\eqref{reorg2}. The large logarithms present there are of the type $\ln X^2Q^2$ or $\ln Y^2Q^2$. Typically the distance between the emitter and the emission point of the gluon, $|X|$ (and $|Y|$) is of order of the average distance between the charges in the projectile wave function. This in turn is precisely the scale $Q_P$ that we have introduced previously. Thus clearly if we choose $Q=Q_P$, no large transverse logarithms will be present in the Hamiltonian. Notice that the
scale $Q_P$ is dynamical, that is it evolves with rapidity. 

 We thus conclude that we should evolve from $Q=Q_T$ down to $Q=Q_P$. In fact, we can do a little better than that. We can in principle, choose the interval of the evolution for  $\mathbb{S}_Q(\mathbf{z})$ to depend on $X$ and $Y$. This makes sense, especially for points $\mathbf{z}$ which are significantly closer to some valence charges than the average distance. Again, examining Eqs.~\eqref{reorg1} and \eqref{reorg2} we see that the optimal choice for the end point of the evolution is simply $Q(\mathbf{z})=\tilde Q$. This choice is naturally of order $Q_P$ for most of the emissions but avoids generating spurious logarithms  for emission points too close to one of
 the emitters.

Thus the resummation of DGLAP logs is achieved by the following straightforward procedure. We solve the RG equation \eqref{dglaplike} for $\mathbb{S}(\mathbf{z})$ with the initial condition Eq.~\eqref{init}  to find $\mathbb{S}_{\tilde Q}(\mathbf{z})$  (for fixed $\mathbf{x}, \mathbf{y}$). We then substitute the solution into Eq.~\eqref{reorg1},\eqref{reorg2}. With $\mathbb{S}_{\tilde Q}$ expressed in terms of $S$ via the solution of RG equations, this yields the resummed Hamiltonian free of large logarithms. Keeping all terms in Eqs.~\eqref{reorg1} and \eqref{reorg2} gives, in addition to the resummation, some genuine small $O(\alpha_s^2)$ corrections. In the following, we will not keep these corrections for simplicity, which is then equivalent to only working with the (running coupling) LO Hamiltonian where $S(\mathbf{z})$ is substituted by $\mathbb{S}_{\tilde Q}(\mathbf{z})$.

The solution of the RG equation Eq.~\eqref{dglaplike} in general is not trivial, but under reasonable assumptions, the equation can be solved analytically in two limits: the dilute limit and the saturation limit.

\subsection{Resummation in the dilute limit - the BFKL.}
Let us first consider the situation where the projectile and the target are both dilute. In this case, the $S$ matrix is close to unity
\begin{equation}
{S}(\mathbf{x})=1+iT^a\alpha^a(\mathbf{x})\,,\quad\quad\quad
\mathbb{S}_Q(\mathbf{x})=1+iT^a\alpha^a_Q(\mathbf{x})\,.
\end{equation}
To linear order in $\alpha_Q$ the RG equation becomes
\begin{align}
   \frac{\partial}{\partial \ln Q^{2}} \alpha_Q^{c}(\mathbf{z}) = 
    \frac{\alpha_s \beta^g_0} {8\pi^2} \, \int d\phi
        \left( 
         \alpha_Q^c(\mathbf{z} + \frac{1}{2} Q^{-1} \mathbf{e}_\phi) 
        - \alpha_Q^{c}(\mathbf{z})  \right) \,.
\end{align}

 Fourier transforming this into momentum space, we obtain
\begin{align}
   \frac{\partial}{\partial \ln Q^{2}} \alpha_Q^{c}( \mathbf{p})\, =\, 
    \frac{\alpha_s \beta^g_0} {8\pi^2}  \, \int d\phi
        \left( e^{i \frac{1}{2} Q^{-1} \mathbf{p} \cdot \mathbf{e}_\phi} 
        - 1 \right)\alpha_Q^{c}(\mathbf{p}) = R(p,Q)\,\alpha_Q^{c}(\mathbf{p})   
\end{align}
with
\begin{equation}\label{rqp}
    R(p,Q) =  \frac{\alpha_s \beta^g_0} {4\pi} \,
    \,\left[J_0\left(\frac{p}{2Q}\right)\,-\,1\right]\,.
    \end{equation}
The solution is
\begin{equation}\label{lin}
  \alpha_Q^{c}(\mathbf{p}) \,=\, \exp\left[-\int_{Q}^{Q_T} \frac{dQ^2}{Q^2} R(p,Q)\right]\, \alpha^{c}( \mathbf{p}) \,.
  \end{equation}

The result of the integral is qualitatively easy to understand. 
Let us rewrite the exponential factor in Eq.~\eqref{lin} in the form
\begin{equation}\label{erqp}
\int_Q^{Q_T} \frac{dQ^2}{Q^2}R(Q,p)=\frac{\alpha_s\beta^g_0}{4\pi^2}\int \limits_{|\mathbf{x}_Q|=\frac{1}{2Q_T}}^{|\mathbf{x}_Q|=\frac{1}{2Q}} \frac{d^2 \mathbf{x}_Q}{\mathbf{x}^2_Q}\left(e^{i\mathbf{p}\cdot\mathbf{x}_Q}-1\right)
\end{equation}
where we introduced a two dimensional vector $\mathbf{x}_Q\equiv \frac{1}{2Q}\mathbf{e}_\phi$.

For $p$ very small $p\ll  Q$, the two terms in Eq.~\eqref{erqp} cancel, and the integral vanishes.  
For $p$ very large, $p\gg  Q_T$ the first term vanishes upon integration. Thus for such $p$ the integral is given by the second term 
\begin{equation}
-\frac{\alpha_s\beta^g_0}{4\pi}\ln\frac{Q_T^2}{Q^2}\,.
\end{equation}
For intermediate values  $2 Q<p< 2Q_T$ the two terms essentially cancel in the range $|\mathbf{x_Q}|<1/p$, whereas for $|\mathbf{x_Q}|>1/p$ only the unity contributes. A good approximation to the result is then 
\begin{equation}
-\frac{\alpha_s\beta^g_0}{4\pi}\ln\frac{p^2}{4Q^2}\,.
\end{equation}
With the logarithmic accuracy, the factor of 4 can be ignored under the logarithm, which we will consistently do in what follows.

Thus  qualitatively, we have 
\begin{eqnarray}\label{qrfixed}
\int_Q^{Q_T} \frac{dQ^2}{Q^2}R(Q,p)\approx
 -\frac{\alpha_s\beta^g_0}{4\pi}\left[\ln\frac{p^2}{Q^2}\theta(Q_T^2-p^2)\theta(p^2-Q^2)+\ln\frac{Q_T^2}{Q^2} \theta(p^2-Q_T^2)\right]\,.
\end{eqnarray}
In momentum space the solution for the field $\alpha_Q$ is approximated by
\begin{align}\label{momalpha}
    \notag 
    \alpha^a_Q(\mathbf {p}) &=\left[\theta(Q^2-p^2)+\left(\frac{p^2}{Q^2}\right)^{\frac{\alpha_s\beta^g_0}{4\pi}}\theta(Q_T^2-p^2)\theta(p^2-Q^2)+\left(\frac{Q_T^2}{Q^2}\right)^{\frac{\alpha_s\beta^g_0}{4\pi}}\theta(p^2-Q_T^2)\right]
    \\ & \times 
    \alpha^a(\mathbf{p})\,.
\end{align}
The above is obtained at a fixed coupling constant. At higher orders the coupling constant will run, and it is clear that in the DGLAP part the running coupling should be taken at the scale $Q$
\begin{equation}
    \alpha_s(Q)=\frac{4\pi}{\beta^g_0}\frac{1}{\log\frac{Q^2}{\Lambda_{\rm QCD}^2}}\,.
\end{equation}
The RHS of Eq.~\eqref{qrfixed} is then modified to 
\begin{eqnarray}\label{qrrunning}
 &&\approx
 -\left[\ln\frac{\ln\frac{p^2}{\Lambda_{\rm QCD}^2}}{\ln\frac{Q^2}{\Lambda_{\rm QCD}^2}}\theta(Q_T^2-p^2)\theta(p^2-Q^2)+\ln\frac{\ln\frac{Q_T^2}{\Lambda_{\rm QCD}^2}}{\ln\frac{Q^2}{\Lambda_{\rm QCD}^2}}\theta(p^2-Q_T^2)\right]\,.
\end{eqnarray}
This modification is formally of the higher order in $\alpha_s$ and we will not consider it further here. We only note that including running $\alpha_S$ as in eq.\eqref{qrrunning} does not present any fundamental difficulties.

Since the JIMWLK Hamiltonian is customarily written in the coordinate space, we need to Fourier transform Eq.~\eqref{momalpha} back into coordinate space. To that end we need
to compute 2-d Fourier transform 
\begin{align}\label{G}
G_Q(\mathbf{x})&\equiv \int \frac{d^2 p}{(2\pi)^2} e^{i \mathbf{p}\cdot\mathbf{x} } 
\int_Q^{Q_T} \frac{dQ^2}{Q^2}R(Q,p)
\simeq 
\int \frac{d^2 p}{(2\pi)^2} e^{i \mathbf{p}\cdot\mathbf{x} }  \\ 
& \times \left[\theta(Q^2-p^2)+\left(\frac{p^2}{Q^2}\right)^{\frac{\alpha_s\beta^g_0}{4 \pi}}\theta(Q_T^2-p^2)\theta(p^2-Q^2)+\left(\frac{Q_T^2}{Q^2}\right)^{\frac{\alpha_s\beta^g_0}{4\pi}}\theta(p^2-Q_T^2)\right]\,.\notag
\end{align}
This is relatively straightforward. The Fourier transform can be calculated exactly in terms of a hyper geometric function. However, given the fact that our calculation in the momentum space itself was approximate, it is much more instructive to write the result in the approximate form, which however exhibits all the important properties of the solution. The details of this derivation are present in Appendix B.
The final result is (we partially expand to order $\alpha_s$, but keep all powers of $\alpha_s\ln\frac{Q_T^2}{Q^2}$)
\begin{itemize}

\item For small $|\mathbf{x}|$: $|\mathbf{x}|<\frac{1}{Q_T}$
\begin{align}\label{sx}
   G_Q(\mathbf{x})=  \left(\frac{Q_T^2}{Q^2}\right)^{\frac{\alpha_s\beta^g_0} {4\pi}}\delta^{(2)}(\mathbf{x}) 
       + {\frac{\alpha_s\beta^g_0}{4\pi}}   \frac{ Q^2}{\pi}\left[1-\left(\frac{Q_T^2}{Q^2}\right)^{1+\frac{\alpha_s\beta^g_0} {4\pi}}\right]\,,
\end{align}
\item For intermediate $|\mathbf{x}|$: $\frac{1}{Q_T}<|\mathbf{x}|<\frac{1}{Q}$
\begin{align}\label{ix}
G_Q(\mathbf{x})=\frac{\alpha_s\beta_0}{4\pi}  \frac{ Q^2}{\pi} \left[1 -    \left(\frac{1}{Q^2x^2}\right) ^{ 1+{\frac{\alpha_s\beta^g_0}{4\pi}} }\right]\,,
\end{align}

\item At large $|\mathbf{x}|$, $|\mathbf{x}|  > \frac1Q$
 the function behaves as 
\begin{align}
\label{lx}
  G_Q(\mathbf{x}) \propto \frac{\alpha_s\beta^g_0}{4\pi}\frac{1 }{(Q|\mathbf{x}|)^{5/2}} \to 0 \,.
\end{align}
\end{itemize}

One can check that the function $G_Q(\mathbf{x})$ satisfies (up to corrections of order $O(\alpha_s^2)$) the ``sum rule'' $\int d^2\mathbf x \, G_Q(\mathbf{x})=1$ as required by Eq.~\eqref{G}. An additional consistency check is that $G_Q(\mathbf{x})\to \delta^2(\mathbf{x})$ when either $\alpha_s=0$ or $Q=Q_T$.

We can now write the solution for $\alpha_Q$ as
\begin{eqnarray}\label{alphar}
\alpha_Q(\mathbf{z})&=&\int d^2\mathbf{x}\,G_Q(\mathbf{z}-\mathbf{x})\alpha(\mathbf{x})\\
&\approx& \left(\frac{Q_T^2}{Q^2}\right)^{\frac{\alpha_s\beta^g_0} {4\pi}}\alpha(\mathbf{z})+\int \limits_{1/Q_T<|\mathbf{x}-\mathbf{z}|<1/Q}d^2\mathbf{x}
\frac{\alpha_s\beta^g_0}{4\pi}  \frac{ Q^2}{\pi} \left[1 -    \left(\frac{1}{Q^2(\mathbf{x}-\mathbf{z})^2}\right) ^{ 1+{\frac{\alpha_s\beta^g_0}{4\pi}} }\right]\alpha(\mathbf {x})\nonumber \,.
\end{eqnarray}
Here in the second line, we  neglected terms of order $\alpha_s$ not enhanced by transverse logarithms. Note, that we keep the second term in the second line, even though formally it has an explicit prefactor of $\alpha_s$. The reason is that the integrand is a very slowly decreasing function of $|\mathbf{x}-\mathbf{z}|$ so that for slowly varying $\alpha(\mathbf{x})$ it integrates to a quantity of order unity. In particular, this term is crucial to satisfy the sum rule $\int d^2\mathbf{z} \, G_Q(\mathbf{z})=1$.

To write the resummed Hamiltonian in the dilute limit we need only to keep the LO term, since the NLO terms now become genuinely small corrections. For consistency, in the LO term we have to expand the left and right charge operators as well as the $S$-matrix $S(\mathbf{z})$ in $\alpha(\mathbf{x})$. This procedure is well known and it leads to the BFKL limit. The only modification in our case is that all the factors $\alpha(\mathbf{z})$ have to be substituted by the resummed amplitude $\alpha_{\tilde Q}(\mathbf{z})$.
Thus in the dilute limit the DGLAP resummation results in the following Hamiltonian
\begin{align}\label{resummedbfkl}
    \notag
    H_{\rm BFKL}^{\rm resummed}&=\int_{\mathbf{x,y,z}}\frac{g(X^{-2})g(Y^{-2})}{8\pi^3}\frac{X\cdot Y}{X^{2}Y^{2}}
    \\ & \times  
    \left[\,(\alpha(\mathbf{x})-\alpha_{\tilde Q}(\mathbf{z}))\,(\alpha(\mathbf{y})-\alpha_{\tilde Q}(\mathbf{z}))\right]^{ab}\frac{\delta^2}{\delta\alpha^a(\mathbf{x}) \delta\alpha^b(\mathbf{y})}\,.
\end{align}
with $\alpha_{\tilde Q}(\mathbf{z})$ given by Eq.~\eqref{alphar} with $\tilde Q(\mathbf{z})$ defined in Eq.~\eqref{tildeq}.

\subsection{The saturation regime}

Now let us consider the saturated regime. Here the momentum $Q_T$ of the target plays a dual role. First, it provides the correlation length for the Wilson lines like before. But now it also provides the saturation scale. This means that a Wilson line separated from any  object composed of other Wilson lines by a distance greater than $1/Q_T$, is small. Although this smallness does not hold configuration by configuration, but only in the sense of averaging over the target ensemble, in practice we can neglect terms that involve products of Wilson lines separated by distances larger than $1/Q_T$. 

If we do this, the evolution simplifies greatly. Recall, that we are evolving from $1/|X|\gg Q_T$ to $Q_T$, so that the distance between the Wilson lines in the nonlinear term in the evolution equation \eqref{dglaplike} is always greater than the correlation length, and therefore the quadratic term is much smaller than the linear term. We can therefore neglect the quadratic term in Eq.~\eqref{dglaplike} altogether. Then the equation becomes linear, and is dominated entirely by the virtual term:
\begin{equation}
\frac{\partial \mathbb{ S}_Q(\mathbf{z})}{\partial \ln Q^2} 
=-\frac{\alpha_s\beta^g_0}{4\pi}\,\mathbb{S}_Q(\mathbf{z})
\end{equation}
with the solution
\begin{equation}\label{solsat}
\mathbb{ S}_Q(\mathbf{z})=\left[\frac{Q_T^2}{Q^2}\right]^{\frac{\alpha_s\beta^g_0}{4\pi} }\mathbb{ S}_{Q_T}(\mathbf{z})\approx\left[\frac{Q_T^2}{Q^2}\right]^{\frac{\alpha_s\beta^g_0}{4\pi} } S(\mathbf{z})\,.
\end{equation}

The resummed Hamiltonian in this regime is
\begin{eqnarray}\label{resummedsat}
H^{\rm resummed}_{\rm JIMWLK}
&=&\int_{\mathbf{x,y,z}}\frac{g(X^{-2})g(Y^{-2})}{8\pi^3}\frac{X\cdot Y}{X^{2}Y^{2}}\,\\
&&\times\Bigg\{\, 
J_{L}^{a}(\mathbf{x})\, J_{L}^{a}(\mathbf{y})\,+\, J_{R}^{a}(\mathbf{x})\, J_{R}^{a}(\mathbf{y})
\,\nonumber\\
&&-\,2J_{L}^{a}(\mathbf{x})\, \left[\theta\left(Q_T-\tilde Q(\mathbf{z})\right)\left[\frac{Q_T^2}{\tilde Q^2(\mathbf{z})}\right]^{\frac{\alpha_s\beta^g_0}{4\pi} } +\theta\left(\tilde Q(\mathbf{z})-Q_T\right)\right] S^{ab}(\mathbf{z})\, J_{R}^{b}(\mathbf{y})\,\Bigg\}\nonumber
\end{eqnarray}
with $\tilde Q$ defined as before Eq.~\eqref{tildeq}, $\tilde Q(\mathbf{z})={\rm max}\left\{\frac{1}{X^2},\frac{1}{Y^2}\right\}$.

Equations~\eqref{resummedsat} and \eqref{resummedbfkl} conclude our derivation in the two simple limits. Our remaining task is to include the DGLAP logarithms due to the presence of the quarks and antiquarks. We will do this in the following section. 

\section{Including quarks}
\label{Sec5}

Including quarks into our framework is fairly straightforward. We need to add the process of a gluon splitting into a $q\bar q$ pair, and additionally to allow a quark to split into a quark and a gluon. The first process already appears in the NLO JIMWLK. The second process appear only at NNLO, because at leading order we do not include quarks into the hadronic wave function. They are only generated at NLO via emission from gluons, and thus can split only at the NNL order. Nevertheless this process is important and affects the DGLAP resummation at leading order.

\subsection{Dressed quarks and gluons.}
The bare quark (antiquark) at the transverse point $\mathbf{x}$ scatters with the eikonal scattering matrix $V(\mathbf{x})$ ($V^\dagger(\mathbf{x})$) in the (anti) fundamental representation of the $SU(N_c)$ group. 
As a unitary matrix, it of course, satisfies $V^\dagger V=1$. In addition
the gluon and the quark eikonal factors are related by
\begin{equation}
    S^{ab}(\mathbf{x})=2\, {\rm Tr} [V^\dagger(\mathbf{x})\tau^a V(\mathbf{x})\tau^b]
\end{equation}
where $\tau^a$ are generators of $SU(N_c)$ in the fundamental representation.

An important thing we have to be aware of, is that these simple properties do not hold for the $S$ matrices of dressed quarks and gluons. Thus for the DGLAP dressed quark states, which we will define in a short while analogously to dressed gluon states we have discussed so far
\begin{equation}\label{not}
\mathbb{V^\dagger}_Q(\mathbf{x})\mathbb{V}_Q(\mathbf{x})\ne 1; \ \ \ \ \ \  \mathbb{S}^{ab}_Q(\mathbf{x})\ne 2\, {\rm Tr} [\mathbb{V}^\dagger_Q(\mathbf{x})\tau^a \mathbb{V}(\mathbf{x})_Q\tau^b]\,.
\end{equation}
This is important since the dressed quark $S$-matrix cannot be inferred automatically from the dressed gluon $S$-matrix, and thus can (and does) satisfy an independent DGLAP-like evolution equation.

Keeping this in mind, we can immediately generalize the equation for the dressed gluon scattering matrix. To leading perturbative order it reads
\begin{align}\label{sfat2}
    \mathbb{S}_Q^{ab}(\mathbf{z}) &= 
     [1+\frac{\alpha_s\beta_0}{4\pi}\ln\frac{\mu^2}{Q^2}]S^{ab}(\mathbf{z}) - \frac{\alpha_s\beta_0^g} {4\pi^2N_c} \int_{|Z|<Q^{-1}}\frac{d^2Z}{Z^2}
         D^{ab}(\mathbf{z} +  \frac{Z}{2}, \mathbf{z} -\frac{Z}{2})   
       \\ &- \frac{\alpha_s\beta_0^q} {4\pi^2N_c} \int \limits_{|Z|<Q^{-1}}\frac{d^2Z}{Z^2}
         D^{F\,ab}(\mathbf{z} +  \frac{Z}{2}, \mathbf{z} -\frac{Z}{2})   
         \notag 
\end{align}
where $\beta_0^q=-\frac{2N_F}{3}$ is the quark contribution to the one loop QCD $\beta$-function, and we have defined
\begin{equation}
 {D}^{F\,ab}(\mathbf{z} +  Z/2, \mathbf{z} - Z/2)  =
    2 \, {\rm Tr}[ \tau^aV(\mathbf{z}+Z/2)  \tau^bV^+(\mathbf{z}-Z/2)]\,.
\end{equation}
As before,
we can rewrite the complete NLO JIMWLK Hamiltonian as
\begin{equation}
H_{\rm JIMWLK}=H^{J\mathbb{S}J}_{Q}+H^{J\mathbb{SS}J}_{Q}+H^{\bar q q}_{Q}+\ldots 
\end{equation}
where now
\begin{eqnarray}\label{reorg1q}
\hspace{-2cm}H^{J\mathbb{S}J}_{ Q}
&=&\int_{\mathbf{x,y,z}}\frac{\alpha_{s}X\cdot Y}{2\pi^{2}X^{2}Y^{2}}\,
\Bigg\{\left(1+\frac{\alpha_s\beta_0}{8\pi}\,\left[\,\ln[X^{2}\mu^{2}]\,+\,\ln[Y^{2}\mu^2]\,\right]\right)
\nonumber \\
&&\hspace{2cm}\times\left[\, J_{L}^{a}(\mathbf{x})\, J_{L}^{a}(\mathbf{y})\,+\, J_{R}^{a}(\mathbf{x})\, J_{R}^{a}(\mathbf{y})\,-\,2J_{L}^{a}(\mathbf{x})\, \mathbb{S}_{\bar Q}^{ab}(\mathbf{z})\, J_{R}^{b}(\mathbf{y})\,\right]\nonumber\\
&&\hspace{2cm}+\frac{\alpha_s\beta_0}{8\pi}\Big[(\ln X^2\tilde Q^2+\ln Y^2\tilde Q^2)\, \left[J_{L}^{a}(\mathbf{x})\, J_{L}^{a}(\mathbf{y})\,+\, J_{R}^{a}(\mathbf{x})\, J_{R}^{a}(\mathbf{y})\right]\\
&&\hspace{3cm}-2
(\ln X^2\bar Q^2+\ln Y^2\bar Q^2)J_{L}^{a}(\mathbf{x})\, \mathbb{S}_{\bar Q}^{ab}(\mathbf{z})\, J_{R}^{b}(\mathbf{y})\,\Big]\Bigg\}\nonumber\,.
\end{eqnarray}
\begin{eqnarray}\label{reorg2q}
 &&H^{J\mathbb{SS}J}_Q=\int_{\mathbf{x},\mathbf{y},\mathbf{z}, \mathbf{z}^{\prime}}\hspace{-0.3cm} K_{JSSJ}(\mathbf{x},\mathbf{y},\mathbf{z}, \mathbf{z}^{\prime})J_{L}^{a}(\mathbf{x})\mathbb{D}_{\bar Q}^{ad}(\mathbf{z},\mathbf{z}^{\prime}) J_{R}^{d}(\mathbf{y})
 -\frac{\alpha^2_{s}\beta^g_0}{4\pi^{3}} \int_{\mathbf{x},\mathbf{y},\mathbf{z}}\frac{X\cdot Y}{X^{2}Y^{2}}
 \mathbb{D}^{ab}_{\bar Q}(\mathbf {z})J_L^a(\mathbf{x})J_R^b(\mathbf{y})\nonumber\\
 &&
 -\left[\int_{\mathbf{x},\, \mathbf{y},\mathbf{z}, \mathbf{z}^{\prime}}\, \frac{N_c}{2}K_{JSSJ}(\mathbf{x},\mathbf{y},\mathbf{z}, \mathbf{z}^{\prime})-\frac{\alpha^2_{s}\beta^g_0}{8\pi^{3}}\int_{\mathbf{x},\mathbf{y},\mathbf{z}} 
 \frac{X\cdot Y}{X^{2}Y^{2}}
 \ln\frac{\mu^2}{\tilde Q^2}
\right]\left[J_R^a(\mathbf{x})J_R^a(\mathbf{y})+J_L^a(\mathbf{x})J_L^a(\mathbf{y})\right]\,
\end{eqnarray}
and
\begin{eqnarray}
    &&H^{\bar q q}_Q=\int_{\mathbf{x},\mathbf{y}, \mathbf{z},\, \mathbf{z}^{\prime}}K_{q\bar{q}}(\mathbf{x},\, \mathbf{y},\mathbf{z}, \mathbf{z}^{\prime}) J_{L}^{a}(\mathbf{x})
    \mathbb{D}^{F\,ab}_{\bar Q}(
    \mathbf{z},\mathbf{z}^{\prime}) J_{R}^{b}(\mathbf{y})-\frac{\alpha^2_{s}\beta^q_0}{4\pi^{3}}\int_{\mathbf{x},\mathbf{y},\mathbf{z}}\frac{X\cdot Y}{X^{2}Y^{2}}
 \mathbb{D}^{F\,ab}_{\bar Q}(\mathbf {z})J_L^a(\mathbf{x})J_R^b(\mathbf{y})\nonumber\\
    &&-\left[\int_{\mathbf{x}, \mathbf{y}, \mathbf{z},\mathbf{z}^{\prime}}\frac{1}{2} K_{q\bar{q}}(\mathbf{x}, \mathbf{y}, \mathbf{z},\mathbf{z}^{\prime})-\frac{\alpha^2_{s}\beta^q_0 } {4\pi^{3}} \int_{\mathbf{x},\mathbf{y},\mathbf{z}}\frac{X\cdot Y}{X^{2}Y^{2}}
    \ln\frac{\mu^2}{\tilde Q^2}
\right]\left[J_R^a(\mathbf{x})J_R^a(\mathbf{y})+J_L^a(\mathbf{x})J_L^a(\mathbf{y})\right]\, .
\end{eqnarray}

For the dressed quark scattering matrix guided by exactly the same considerations, we have at the leading perturbative order
\begin{equation}
    \mathbb{V}_Q^{\alpha\beta}(\mathbf{z})=\left[1+\frac{3}{2}\frac{\alpha_sC_F}{4\pi}\ln\frac{\mu^2}{Q^2}\right]V^{\alpha\beta}(\mathbf{z})- \frac{3}{2}\frac{\alpha_s} {4\pi^2} \int \limits_{|Z|<Q^{-1}}\frac{d^2Z}{Z^2}\left[\tau^a
         S^{ab}(\mathbf{z}+Z/2)V(\mathbf{z}-Z/2)\tau^b\right]^{\alpha\beta}
\end{equation}
where we  have used the fact that for the quark-gluon splitting function
\begin{equation}
    \int_0^1d\xi\left[\frac{1+(1-\xi)^2}{\xi}\right]_+=-\frac{3}{2}
\end{equation}
and
\begin{equation}
    C_F=\frac{N_c^2-1}{2N_c}
    \,.
\end{equation}
Given these perturbative expressions we can write down the DGLAP-like equations which now include quarks.
For the adjoint scattering matrix we have
\begin{eqnarray}\label{dglaplikeg}
&&   \frac{\partial}{\partial \ln Q^2} \mathbb{S}_Q^{ab}(\mathbf{z}) = -
    \frac{\alpha_s}{4\pi} 
    \Bigg[\beta_0 \mathbb{S}_Q^{ab}(\mathbf{z})\\
    &&\quad\quad-\frac{1}{N_c}\int \frac{d\phi}{2\pi}
       \left(\beta_0^g \mathbb{D}_Q^{ab}(\mathbf{z} + \frac{1}{2} Q^{-1} \mathbf{e}_\phi, \mathbf{z} -\frac{1}{2} Q^{-1} \mathbf{e}_\phi)  +\beta_0^q \mathbb{D}_Q^{F\,ab}(\mathbf{z} + \frac{1}{2} Q^{-1} \mathbf{e}_\phi, \mathbf{z} -\frac{1}{2} Q^{-1} \mathbf{e}_\phi) \right) \Bigg]\nonumber
\end{eqnarray}
and for the fundamental scattering matrix
\begin{equation}\label{dglaplikeq}
     \frac{\partial}{\partial \ln Q^2} \mathbb{V}_Q^{\alpha\beta}(\mathbf{z}) = -\frac{3}{2}
    \frac{\alpha_s}{2\pi} 
    \Bigg[ C_F\mathbb{V}_Q^{\alpha\beta}(\mathbf{z})-\int \frac{d\phi}{2\pi}
        \mathbb{D}_Q^{(FA)\alpha\beta}(\mathbf{z} + \frac{1}{2} Q^{-1} \mathbf{e}_\phi, \mathbf{z} -\frac{1}{2} Q^{-1} \mathbf{e}_\phi)\Bigg]
\end{equation}
where 
\begin{eqnarray}
   && \mathbb{D}^{F\,ab}_Q(\mathbf{z} +  Z/2, \mathbf{z} - Z/2)  \equiv
    2 {\rm Tr}\, [ \tau^a  \mathbb{V}_Q(\mathbf{z}+Z/2)  \tau^b \mathbb{V}_Q^+(\mathbf{z}-Z/2)]\,, \nonumber\\
     && \mathbb{D}_Q^{(FA)\alpha\beta}(\mathbf{z}+Z/2,\mathbf{z}-Z/2)\equiv \left[\tau^a
         \mathbb{S}^{ab}_Q(\mathbf{z}+Z/2)\mathbb{V}_Q(\mathbf{z}-Z/2)\tau^b\right]^{\alpha\beta}\,.
\end{eqnarray}
The interesting albeit not unexpected property of these equations is that the evolution of the adjoint and fundamental scattering matrices mix. This means that, although we only need to find the adjoint $\mathbb{S}_Q$ in order to perform the DGLAP resummation of the JIMWLK Hamiltonian at NLO, we have to consider also the $Q^2$ evolution of $\mathbb{V}_Q$. This is so even though the terms in the JIMWLK Hamiltonian in which we would have to use the solution of eq.\eqref{dglaplikeq} appear only starting at NNLO.

\subsection{The resummation - saturated regime}

As just mentioned, to resum the transverse logarithms at leading order we do not need to know $\mathbb{V}_Q$, but only $\mathbb{S}_Q$. This makes the resummation in saturation regime very straightforward. In Eq.~\eqref{dglaplikeg} we can neglect, as before the ``real'' $\mathbb{D}_Q$ terms which involve the product of two scattering matrices. The solution is then exactly the same as Eq.~\eqref{solsat} but with the complete one loop $\beta$-function
\begin{equation}\label{solsatg}
\mathbb{ S}_Q(\mathbf{z})=\left[\frac{Q_T^2}{Q^2}\right]^{\frac{\alpha_s\beta_0}{4\pi} }\mathbb{ S}_{Q_T}(\mathbf{z})\approx\left[\frac{Q_T^2}{Q^2}\right]^{\frac{\alpha_s\beta_0}{4\pi} } S(\mathbf{z})\,.
\end{equation}
For completeness, we can also straightforwardly solve for $\mathbb{V}_Q$, even though it is not needed to obtain the resummed Hamiltonian
\begin{equation}\label{solsatq}
\mathbb{ V}_Q(\mathbf{z})=\left[\frac{Q_T^2}{Q^2}\right]^{\frac{3\alpha_s(N^2_c-1)}{8\pi N_c} }\mathbb{ V}_{Q_T}(\mathbf{z})\approx\left[\frac{Q_T^2}{Q^2}\right]^{\frac{3\alpha_s(N^2_c-1)}{8\pi N_c} } V(\mathbf{z})\,.
\end{equation}
The resummed JIMWLK Hamiltonian is then

\begin{eqnarray}\label{resummedsatg}
H^{\rm resummed}_{\rm JIMWLK}
&=&\int_{\mathbf{x,y,z}}\frac{g(X^{-2})g(Y^{-2})}{8\pi^3}\frac{X\cdot Y}{X^{2}Y^{2}}\,\Bigg\{\, 
J_{L}^{a}(\mathbf{x})\, J_{L}^{a}(\mathbf{y})\,+\, J_{R}^{a}(\mathbf{x})\, J_{R}^{a}(\mathbf{y})
\\
&-&2J_{L}^{a}(\mathbf{x})\, \left[\theta\left(Q_T-\tilde Q(\mathbf{z})\right)\left[\frac{Q_T^2}{\tilde Q^2(\mathbf{z})}\right]^{\frac{\alpha_s\beta_0}{4\pi} } +\theta\left(\tilde Q(\mathbf{z})-Q_T\right)\right] S^{ab}(\mathbf{z})\, J_{R}^{b}(\mathbf{y})\,\Bigg\}\nonumber \,.
\end{eqnarray}

If we wanted to keep nonlogarithmic $O(\alpha_s^2)$ terms in the resummed Hamiltonian, we would need to use the solutions Eqs.~\eqref{solsatg}, \eqref{solsatq} and substitute them into the $O(\alpha_s^2)$ terms in the Hamiltonian, but we will not concern ourselves with this here.

\subsection{The resummation: the dilute regime}
Here we again assume that both, $\mathbb{S}_Q$ and $\mathbb{V}_Q$ are close to unity:
\begin{equation}
    \mathbb{S}_Q(\mathbf{x})=1+iT^a\alpha^a_Q(\mathbf{x}); \ \ \ \ \ \ \mathbb{V}_Q(\mathbf{x})=1+i\tau^a\alpha^{Fa}_Q(\mathbf{x})\,.
\end{equation}
Note, that due to Eq.~\eqref{not}, we are not assuming that $\alpha_Q$ is equal to $\alpha^F_Q$.
The evolution equation now becomes a little more complicated, since it mixes $\alpha_Q$ with $\alpha^F_Q$.

The equations for the $Q^2$ evolution now are
\begin{align}
   \frac{\partial}{\partial \ln Q^{2}} \alpha_Q^{c}(\mathbf{z}) = 
    \frac{\alpha_s \beta^g_0} {8\pi^2} \, \int d\phi
        \left( 
         \alpha_Q^c(\mathbf{z} + \frac{1}{2} Q^{-1} \mathbf{e}_\phi) 
        - \alpha_Q^{c}(\mathbf{z})  \right) \notag\\
        +\frac{\alpha_s \beta^q_0} {8\pi^2} \, \int d\phi
        \left( 
         \alpha_Q^{Fc}(\mathbf{z} + \frac{1}{2} Q^{-1} \mathbf{e}_\phi) 
        - \alpha_Q^{c}(\mathbf{z})  \right)
\end{align}
and
\begin{align}
   \frac{\partial}{\partial \ln Q^{2}} \alpha_Q^{Fc}(\mathbf{z}) = 
   \frac{3}{2} \frac{\alpha_s } {4\pi^2} \, \int d\phi &
        \left( \frac{N_c}{2}
         \alpha_Q^{c}(\mathbf{z} + \frac{1}{2} Q^{-1} \mathbf{e}_\phi) -\frac{1}{2N_c}\alpha_Q^{Fc}(\mathbf{z} + \frac{1}{2} Q^{-1} \mathbf{e}_\phi)\right. 
         \left. 
        - C_F \, 
        \alpha_Q^{Fc}(\mathbf{z})  \right)\,.
        \end{align}
To solve these equations we follow the same route as in the previous section, i.e. transform into momentum space. Introducing the function
\begin{equation}\label{rqpq}
    \tilde R(p,Q) =  \,
     J_0\left(\frac{p}{2Q}\right)\,-\,1
    \end{equation}
we obtain local equations in momentum space
\begin{equation}
   \frac{\partial}{\partial \ln Q^{2}} \alpha_Q^{c}(\mathbf{p}) = 
    \frac{\alpha_s \beta^g_0} {4\pi} \, \tilde R(p,Q)\alpha_Q^c(\mathbf{p})+\frac{\alpha_s \beta^q_0} {4\pi}\left\{\left[\tilde R(p, Q)+1\right]\alpha^{Fc}_Q(\mathbf{p})-\alpha^c_Q(\mathbf{p})\right\}\,,
\end{equation}
\begin{equation}
   \frac{\partial}{\partial \ln Q^{2}} \alpha_Q^{Fc}(\mathbf{p}) = 
   \frac{3}{2} \frac{\alpha_s } {2\pi} \,\left\{\left[ \tilde R(p, Q)+1\right]
        \left( \frac{N_c}{2}
         \alpha_Q^{c}(\mathbf{p}) -\frac{1}{2N_c}\alpha_Q^{Fc}(\mathbf{p})\right)
        - C_F\alpha_Q^{Fc}(\mathbf{p}) \right\}\,.
        \end{equation}
These are still pretty complicated. We now however employ the same simplification we used to analyze the no-quark case. Our calculation in Eq.~\eqref{qrfixed} amounts to the following simple approximation for $\tilde R$
\begin{equation}
    \tilde R(p,Q)\approx 0; \ \ p<Q;\ \ \ \ \tilde R(p,Q)=-1;\ \ p>Q\,.
\end{equation}
Then for $p<Q$ we have
\begin{eqnarray}
   \frac{\partial}{\partial \ln Q^{2}} \alpha_Q^{c}(\mathbf{p})& =& 
    \frac{\alpha_s \beta^q_0} {4\pi}\left[\alpha^{Fc}_Q(\mathbf{p})-\alpha^c_Q(\mathbf{p})\right]\,,\nonumber\\
   \frac{\partial}{\partial \ln Q^{2}} \alpha_Q^{Fc}(\mathbf{p})& =& 
   \frac{3}{2} \frac{\alpha_sN_c } {4\pi} \,\left[
         \alpha_Q^{c}(\mathbf{p}) -\alpha_Q^{Fc}(\mathbf{p})
        \right]\,.
        \end{eqnarray}
While for $p>Q$
\begin{eqnarray}\label{p>}
   \frac{\partial}{\partial \ln Q^{2}} \alpha_Q^{c}(\mathbf{p}) &=& 
   - \frac{\alpha_s } {4\pi}\beta_0\alpha_Q^c(\mathbf{p})\,,\nonumber\\
   \frac{\partial}{\partial \ln Q^{2}} \alpha_Q^{Fc}(\mathbf{p}) &=& 
   -\frac{3}{2} \frac{\alpha_s } {2\pi} \,
        C_F\alpha_Q^{Fc}(\mathbf{p}) \,.
        \end{eqnarray}
        These equations are easily solved.
Recall that the interval of the evolution is from $\tilde Q$ to $Q_T$. Thus for
$p<\tilde Q$ we only need to solve the first set of equations. Their diagonalized form is
\begin{eqnarray}\label{p<}
    \frac{\partial}{\partial \ln Q^{2}} \left[\alpha_Q^{c}(\mathbf{p})-\alpha_Q^{Fc}(\mathbf{p})\right]&=&-\frac{\alpha_s}{4\pi}\left(\beta_0^q+\frac{3N_c}{2}\right)\left[\alpha_Q^{c}(\mathbf{p})-\alpha_Q^{Fc}(\mathbf{p})\right]\,,\nonumber\\
    \frac{\partial}{\partial \ln Q^{2}} \left[\frac{3N_c}{2}\alpha_Q^{c}(\mathbf{p})+\beta_0^q\alpha_Q^{Fc}(\mathbf{p})\right]&=&0\,.
\end{eqnarray}
Our initial condition is $\alpha_{Q_T}=\alpha^F_{Q_T}=\alpha$. The solution of Eq.~\eqref{p<} with this initial condition is simply
\begin{equation}
    \alpha_{Q}(p<Q)=\alpha^F_{Q}(p<Q)=\alpha(p<Q)\,.
\end{equation}
It is equally straightforward to solve for $p>Q_T$. For that, we need to solve Eq.~\eqref{p>}, which is straightforward. The solution is
\begin{equation}\label{plarge}
    \alpha^c_Q(p>Q_T)=\left[\frac{Q_T^2}{Q^2}\right]^{\frac{\alpha_s}{4\pi}\beta_0}\alpha^c(p>Q_T);\ \ \ \ \ \ \alpha^{Fc}_Q(p>Q_T)=\left[\frac{Q_T^2}{Q^2}\right]^{\frac{\alpha_s}{4\pi}3C_F}\alpha^c(p>Q_T)\,.
\end{equation}
Finally, we consider momenta $Q_T>p>Q$. For these momenta while evolving from $Q$ to $Q_T$, we should evolve $\alpha$ according to Eq.~\eqref{p>} between $Q$ and $p$, and then according to Eq.~\eqref{p<} between $p$ and $Q_T$. The first part of the evolution results in the same expressions as Eq.~\eqref{plarge} but with $Q_T$ substituted by $p$
\begin{eqnarray}\label{plarge1}
   && \tilde\alpha^c_Q(Q_T>p>Q)=\left[\frac{p^2}{Q^2}\right]^{\frac{\alpha_s}{4\pi}\beta_0}\alpha^c(Q_T>p>Q);\\
   && \tilde\alpha^{Fc}_Q(Q_T>p>Q)=\left[\frac{p^2}{Q^2}\right]^{\frac{\alpha_s}{4\pi}3C_F}\alpha^c(Q_T>p>Q)\nonumber
\end{eqnarray}
where $\tilde\alpha$ denotes the solution evolved from $Q$ to $p$. This $\tilde\alpha$ now serves as the initial condition for the evolution from $p$ to $Q_T$ according to Eq.~\eqref{p<}.
Eventually, we find the solution: 
\begin{eqnarray}
    \alpha^c_Q(p)-\alpha^{Fc}_Q(p)&=&\left[\frac{p^2}{Q_T^2}\right]^{\frac{\alpha_s}{4\pi}\left[\beta_0^q+\frac{3N_c}{2}\right]}\left\{\left[\frac{p^2}{Q^2}\right]^{\frac{\alpha_s}{4\pi}\beta_0}-\left[\frac{p^2}{Q^2}\right]^{\frac{\alpha_s}{4\pi}3C_F}\right\}\alpha^c(p)\nonumber\\
     \frac{3N_c}{2}\alpha^c_Q(p)+\beta_0^q\alpha^{Fc}_Q(p)&=&\left\{\frac{3N_c}{2}\left[\frac{p^2}{Q^2}\right]^{\frac{\alpha_s}{4\pi}\beta_0}+\beta_0^q\left[\frac{p^2}{Q^2}\right]^{\frac{\alpha_s}{4\pi}3C_F}\right\}\alpha^c(p)
\end{eqnarray}
        from which it follows
\begin{eqnarray}
 \alpha^c_Q(p)&=&\frac{1}{\beta_0^q+\frac{3N_c}{2}}   \Bigg\{\left(\frac{3N_c}{2}+\beta_0^q\left[\frac{p^2}{Q_T^2}\right]^{\frac{\alpha_s}{4\pi}\left[\beta_0^q+\frac{3N_c}{2}\right]}\right)\left[\frac{p^2}{Q^2}\right]^{\frac{\alpha_s}{4\pi}\beta_0}
 \nonumber\\
&&\ \ \ \ \ \ \ \ \ \ \ \ \ \ +\beta_0^q\left(1-\left[\frac{p^2}{Q_T^2}\right]^{\frac{\alpha_s}{4\pi}\left[\beta_0^q+\frac{3N_c}{2}\right]}\right)\left[\frac{p^2}{Q^2}\right]^{\frac{\alpha_s}{4\pi}3C_F}\Bigg\}\alpha^c(p)\,,\\
\alpha^{Fc}_Q(p)&=&\frac{1}{\beta_0^q+\frac{3N_c}{2}}   \Bigg\{\frac{3N_c}{2}\left(1-\left[\frac{p^2}{Q_T^2}\right]^{\frac{\alpha_s}{4\pi}\left[\beta_0^q+\frac{3N_c}{2}\right]}\right)\left[\frac{p^2}{Q^2}\right]^{\frac{\alpha_s}{4\pi}\beta_0}
 \nonumber\\
&&\ \ \ \ \ \ \ \ \ \ \ \ \ \ +\left(\beta_0^q+\frac{3N_c}{2}\left[\frac{p^2}{Q_T^2}\right]^{\frac{\alpha_s}{4\pi}\left[\beta_0^q+\frac{3N_c}{2}\right]}\right)\left[\frac{p^2}{Q^2}\right]^{\frac{\alpha_s}{4\pi}3C_F}\Bigg\}\alpha^c(p)\,.
\end{eqnarray}
The last expression we present for completeness, since as mentioned above we only need $\alpha_Q$ to perform the resummation in JIMWLK at leading order.
Thus for all values of momentum, we find
\begin{eqnarray}
\label{Eq:alphac}
 \alpha^c_Q(\mathbf{p})&=&\alpha^c(\mathbf{p})\theta(Q-p)\\
 &+&\frac{1}{\beta_0^q+\frac{3N_c}{2}}   \Bigg\{\left(\frac{3N_c}{2}+\beta_0^q\left[\frac{p^2}{Q_T^2}\right]^{\frac{\alpha_s}{4\pi}\left[\beta_0^q+\frac{3N_c}{2}\right]}\right)\left[\frac{p^2}{Q^2}\right]^{\frac{\alpha_s}{4\pi}\beta_0}
 \nonumber\\
&&\ \ \ \ \ \ \ \ \ \ \ \ \ \ +\beta_0^q\left(1-\left[\frac{p^2}{Q_T^2}\right]^{\frac{\alpha_s}{4\pi}\left[\beta_0^q+\frac{3N_c}{2}\right]}\right)\left[\frac{p^2}{Q^2}\right]^{\frac{\alpha_s}{4\pi}3C_F}\Bigg\}\alpha^c(\mathbf{p})\theta(p-Q)\theta(Q_T-p)\nonumber\\
&+&\left[\frac{Q_T^2}{Q^2}\right]^{\frac{\alpha_s}{4\pi}\beta_0}\alpha^c(\mathbf{p})\theta(p-Q_T)\,,\nonumber\\
\label{Eq:alphaFc}
\alpha^{Fc}_Q(\mathbf{p})&=&\alpha^c(\mathbf{p})\theta(Q-p)\\
&+&\frac{1}{\beta_0^q+\frac{3N_c}{2}}   \Bigg\{\frac{3N_c}{2}\left(1-\left[\frac{p^2}{Q_T^2}\right]^{\frac{\alpha_s}{4\pi}\left[\beta_0^q+\frac{3N_c}{2}\right]}\right)\left[\frac{p^2}{Q^2}\right]^{\frac{\alpha_s}{4\pi}\beta_0}
 \nonumber\\
&&\ \ \ \ \ \ \ \ \ \ \ \ \ \ +\left(\beta_0^q+\frac{3N_c}{2}\left[\frac{p^2}{Q_T^2}\right]^{\frac{\alpha_s}{4\pi}\left[\beta_0^q+\frac{3N_c}{2}\right]}\right)\left[\frac{p^2}{Q^2}\right]^{\frac{\alpha_s}{4\pi}3C_F}\Bigg\}\alpha^c(\mathbf{p})\theta(p-Q)\theta(Q_T-p)\nonumber\\
&+&\left[\frac{Q_T^2}{Q^2}\right]^{\frac{\alpha_s}{4\pi}3C_F}\alpha^c(\mathbf{p})\theta(p-Q_T)\,.\nonumber
\end{eqnarray}

Fourier transformation of these expressions into coordinate space does not present any additional difficulties and can be performed in the same way as in the previous section.
We obtain 
\begin{eqnarray}\label{alpharFull}
\alpha_Q(\mathbf{z})\,
&&\approx\, \left(\frac{Q_T^2}{Q^2}\right)^{\frac{\alpha_s\beta_0} {4\pi}}\alpha(\mathbf{z}) \\ &&\hspace{-0.5cm}+ \hspace{-0.5cm}
\int \limits_{{1\over Q_T}<|\mathbf{x}-\mathbf{z}|<{1\over Q}}
\frac{d^2\mathbf{x}}{\beta_0^q+\frac{3N_c}{2}}   \Bigg\{\left(\frac{3N_c}{2}+\beta_0^q Q_T^2\left[\frac{1}{(\mathbf{x}-\mathbf{z})^2 Q_T^2}\right]^{\frac{\alpha_s}{4\pi}\left[\beta_0^q+\frac{3N_c}{2}\right]+1}\right) Q^2\left[\frac{1}{(\mathbf{x}-\mathbf{z})^2 Q^2}\right]^{\frac{\alpha_s}{4\pi}\beta_0+1}
 \nonumber\\
&&\ \ \ \ \ \ \ \ \ \ \ +\beta_0^q\left(1-Q_T^2 \left[\frac{1}{(\mathbf{x}-\mathbf{z})^2 Q_T^2}\right]^{\frac{\alpha_s}{4\pi}\left[\beta_0^q+\frac{3N_c}{2}\right]+1}\right)Q^2 \left[\frac{1}{(\mathbf{x}-\mathbf{z})^2 Q^2}\right]^{\frac{\alpha_s}{4\pi}3C_F+1}\Bigg\}\,\alpha(\mathbf{x}) 
\notag 
\end{eqnarray}
and 
\begin{eqnarray}\label{alpharFFull}
\alpha^{F}_Q(\mathbf{z})\,
&&\approx \,\left(\frac{Q_T^2}{Q^2}\right)^{^{\frac{\alpha_s}{4\pi}3C_F} }\alpha^{F}(\mathbf{z}) \\ &&\hspace{-0.9cm} + \hspace{-0.5cm} 
\int \limits_{{1\over Q_T}<|\mathbf{x}-\mathbf{z}|<{1\over Q}}
\frac{d^2\mathbf{x}}{\beta_0^q+\frac{3N_c}{2}}   \Bigg\{\left(\beta_0^q  + \frac{3N_c}{2}Q_T^2\left[\frac{1}{(\mathbf{x}-\mathbf{z})^2 Q_T^2}\right]^{\frac{\alpha_s}{4\pi}\left[\beta_0^q+\frac{3N_c}{2}\right]+1}\right) Q^2\left[\frac{1}{(\mathbf{x}-\mathbf{z})^2 Q^2}\right]^{\frac{3\alpha_s}{4\pi}C_F+1}
 \nonumber\\
&&\ \ \ \ \ \  \ \ \ + \,\frac{3 N_c}2 \left(1-Q_T^2 \left[\frac{1}{(\mathbf{x}-\mathbf{z})^2 Q_T^2}\right]^{\frac{\alpha_s}{4\pi}\left[\beta_0^q+\frac{3N_c}{2}\right]+1}\right)Q^2 \left[\frac{1}{(\mathbf{x}-\mathbf{z})^2 Q^2}\right]^{\frac{\alpha_s}{4\pi}\beta_0+1}\Bigg\}\,\alpha^{F}(\mathbf{x}) \,.
\notag 
\end{eqnarray}
The resummed Hamiltonian is now given by Eq.\eqref{resummedbfkl} 
with $\alpha_{\tilde Q}(\mathbf{z})$ given by Eq.~\eqref{alpharFull} with $\tilde Q(\mathbf{z})$ defined in Eq.~\eqref{tildeq}.

\section{Discussion}
\label{Sec6}

In this paper we considered a partial resummation of transverse logarithms in the NLO JIMWLK equation. We have shown that some of these logs, which are proportional to the one-loop coefficient of the QCD $\beta$-function are not associated with the renormalization of the coupling constant, but instead encode DGLAP corrections to the low $x$ evolution.
These DGLAP corrections are large whenever there is a large disparity between the correlation lengths (or saturation momenta) in the projectile and the target. Since this is precisely the regime in which the JIMWLK equation is supposed to be applicable, resumming these corrections is indeed essential in the region of validity of JIMWLK equation.

We showed how to resum these corrections by solving a DGLAP-like equation that describes the scale evolution of the scattering matrix of dressed gluons and quarks. The $S$ matrix needs to be evolved from the scale of the target's correlation length (saturation momentum) to that of the projectile. We solved these equations explicitly in two limiting cases. The first is when the target is dilute, and the correlation length is not associated with the saturation momentum. In this case, both real and virtual corrections are important for the DGLAP-like evolution.  The second case is when the target is saturated with large saturation momentum. In this case, the evolution is dominated by the virtual term and the solution has a  simple scaling form.

The running coupling corrections of course are still there and also need to be resummed. However, in our approach, we do not see a need for any elaborate choice for the scale of the coupling constant. Although, of course, strictly speaking, one needs higher order corrections to set the coupling scale, we observe that the most naive choice, i.e., the distance between the emitter and the emitted gluon, is a perfectly good candidate to set the scale of the running strong charge. Choosing this scale and including the DGLAP resummation takes care of all large logarithms associated with the one loop $\beta$-function in the NLO JIMWLK.

There is one important point we need to realize about the DGLAP-like corrections  discussed here. Although we have been referring throughout this paper to ``dressed gluons'' and ``dressed quarks'', these are actually not the same normalized dressed states one encounters usually in QCD calculations. The dressed parton states are created by DGLAP emission with the complete splitting function. In the context of JIMWLK evolution, the situation is different. The low $x$ part of the splitting function is associated with the JIMWLK evolution in rapidity, rather than the DGLAP splittings. As we  noted earlier, this is the reason we introduced a plus prescription that subtracts the contribution of both poles at $\xi=0$ and $\xi=1$ in Eqs.~\eqref{sfat},~\eqref{xi++} and those following. This subtraction  compensates for the inclusion of the low $x$ part of the splitting function, which the standard low $x$ evolution (be it JIMWLK or BK) uses formally at all values of $x$. As a result, whereas the (real contribution to the) usual splitting function is positive since it has the meaning of emission probability, in our case, the analogous quantity is negative, e.g., the coefficient in front of the second (``real'') term in Eq.~\eqref{sfat2}. 

This negativity has a very stark consequence to our final results. Consider for example, Eq.~\eqref{solsat} which gives the solution for the scattering matrix of our ``dressed gluon'' at resolution scale $Q$. If this were the physical dressed gluon state, one would, of course, expect that for $Q<Q_T$, the spatial extent of this state is large, and the scattering matrix is smaller than $S$ since the scattering amplitude should be larger for a spatially extended state. However Eq.~\eqref{solsat} behaves in exactly the opposite way. As $Q$ gets smaller, $\mathbb{S}_Q$ {\it increases}, i.e. becomes {\it less} saturated. This is precisely the effect of subtracting the low x part from the DGLAP splitting function. For this reason, we  referred to the corrections considered here as ``DGLAP-like" rather than simply ``DGLAP".

The low $x$ evolution, therefore, partitions the DGLAP logarithms in a rather amusing way. In the Hamiltonian itself, only a single gluon emission with  the low $x$ part of the splitting function appears since the Hamiltonian is extracted from terms in the $S$-matrix, which are linear in the longitudinal phase space $\Delta Y$. However, calculating $H_{\rm JIMWLK}$ at higher orders in $\alpha_s$ brings an arbitrary number of emissions with the finite part of the splitting function (with the low x poles subtracted), which look like emissions with {\it negative} probability. These are the logarithms that we showed how to resum in the present paper.
If we were to look at the second order in $\Delta Y$ contribution to the scattering matrix, i.e. the second iteration of the JIMWLK Hamiltonian in the evolution, we would find two eikonal emissions, i.e. two emissions with the low $x$ part of the splitting function. Thus at this order we should be able to reconstruct the full splitting function for two emissions, but the rest would still look like emissions with negative probability. And so it goes. 

By what time do we recover the complete positive probability (splitting function) at all orders in the DGLAP part of the evolution? It looks natural that one would need to evolve in $Y$ by the interval that is of the same order as the interval of the DGLAP evolution, i.e. $\Delta Y\approx \ln\frac{Q^2_T}{Q^2_P}$ to restore the positivity of probabilities for  all splittings. Interestingly, parametrically by the end of this evolution in $Y$ we expect $Q_P^2(Y)\approx Q_T^2(Y)$. In principle, to evolve the system in $Y$ any further, we need to go beyond the JIMWLK approximation since our projectile is not anymore dilute relative to the target. It is not clear to us at this point whether this observation has any significance, apart from strengthening the point that the DGLAP resummations have to be always performed within the JIMWLK evolution.

For the future there are two obvious directions in which one should extend our results. First, it would be interesting to solve the full RG equation beyond the dilute and dense limits. The equation itself has a form somewhat similar to the BK equation, with the obvious difference that the basic object is a matrix rather than a scalar amplitude. It may not be possible to solve this equation analytically, but it is certainly worth a try.

Second, in the present paper we have only dealt with the resummation of DGLAP logarithms at leading order. Correspondingly, we did not keep genuinely perturbative $O(\alpha_s^2)$ corrections in the resummed Hamiltonian. For precision applications one certainly would need to do that. In principle it is a straightforward matter, as one would just need to keep $O(\alpha_s^2)$ in $H_{\rm JIMWLK}$ after substitution of $\mathbb{S}_{\bar Q}$. However, as always in going beyond leading order, one has to be careful about details and implement a consistent factorization scheme throughout the calculation, including more careful solution for $\alpha_Q$ and $\mathbb{S}_Q$. At this level of accuracy one also needs to include the running of the QCD coupling constant in the DGLAP evolution along the lines indicated in Eq.~\eqref{qrrunning}.

In the regime where neither projectile nor target are saturated, the JIMLWK equation reduces to the BFKL equation. Various schemes that include both BFKL and DGLAP logarithms have been considered in the literature 
\cite{Salam:1998tj,Ciafaloni:1999au,Ciafaloni:1999yw,Ciafaloni:2003ek,Ciafaloni:2003rd,Ciafaloni:2003kd,Thorne:2001nr,White:2006yh,Altarelli:1999vw,Altarelli:2000mh,Altarelli:2001ji,Altarelli:2003hk,Altarelli:2008aj,Ball:2017otu}.
None of these approaches however are applied in coordinate space which is natural for the JIMWLK equation. Consequently at this stage we were unable to relate our procedure to those in the above papers.

We note that in recent years attempts have been, and are being made to unify the low and intermediate $x$ physics including the saturation ideas\cite{Jalilian-Marian:2018iui, Boussarie:2020fpb, Boussarie:2021wkn}. Although technically our approach here is very different, the physics problem we address has the same origin and we hope these various approaches will converge in future.

\newpage

\appendix


\section{Appendix: The Right and Left Charges.}
 We may want to express $H_{\rm JIMWLK}$ entirely in terms of the dressed gluons scattering matrix $\mathbb{S}_Q$. This would require also expressing the left - and right rotation operators as operators acting on functions of $\mathbb{S}_Q$ rather than $S$. 
We consider the right charges first.  Recall that
\begin{equation}
J^a_R(\mathbf{x}){S}^{eb}(\mathbf {y})=[S(\mathbf{y})T^a]^{eb}\delta^2(\mathbf{x}-\mathbf{y});\ \ \ \ 
J^a_R(\mathbf{x}){S}^{T eb}(\mathbf {y})=-[T^aS^T(\mathbf{y})]^{eb}\delta^2(\mathbf{x}-\mathbf{y})\,.
\end{equation}
Thus
\begin{equation}
J^a_R(\mathbf{x})=-{\rm Tr}\left[\frac{\partial}{\partial S(\mathbf{x})}T^aS^T(\mathbf{x})\right]   \, . 
\end{equation}

In principle the problem is algebraic. We should just use the chain rule to express the charges in terms of $\mathbb{S}_Q$
\begin{eqnarray}
J^a_R(\mathbf{x})&=&-{\rm Tr}\left[\frac{\partial}{\partial S(\mathbf{x})}T^aS^T(\mathbf{x})\right]=-\int 
_y{\rm Tr}\left[\frac{\partial \mathbb{S}^{eb}_Q(\mathbf{y})}{\partial S(\mathbf{x})}\frac{\partial}{\partial \mathbb{S}^{eb}_Q(\mathbf{y})}T^aS^T(\mathbf{x})\right]\nonumber\\
&=&\int_\mathbf{y}\left[J^a_R(\mathbf{x})\mathbb{S}^{eb}_Q(\mathbf {y})\right]\frac{\partial}{\partial \mathbb{S}^{eb}_Q(\mathbf{y})}\,.
\end{eqnarray}
Using the definition we have
\begin{eqnarray}
\frac{\partial\mathbb{S}_Q^{eb}(\mathbf{y})}{\partial S^{cd}(\mathbf{x})}&=&[1+\frac{\alpha_s\beta_0^g}{4\pi}\ln\frac{\mu^2}{Q^2}]\delta^{ec}\delta^{bd}\delta^2(\mathbf{y}-\mathbf{x})\\
&-&2\frac{\alpha_s\beta_0^g} {4\pi^2 N_c}\frac{1}{(\mathbf{x}-\mathbf{y})^2}\left[T^eS(2\mathbf{y}-\mathbf{x})T^b\right]_{cd}\Theta[(2Q)^{-2}-(\mathbf{x}-\mathbf{y})^2]\,.\nonumber
\end{eqnarray}
So that
\begin{eqnarray}\label{jrq}
 J^a_R(\mathbf{x})&=&
-{\rm Tr}\left[\frac{\partial}{\partial \mathbb{S}_Q(\mathbf{x})}T^a\mathbb{S}^T_Q(\mathbf{x})\right]+\frac{\alpha_s\beta_0^g} {4\pi^2N_c}\int^{\frac{1}{2Q}}_{\mathbf{y-x}}\frac{1}{(\mathbf{x}-\mathbf{y})^2}\nonumber\\
&\times&\Bigg[{\rm Tr}\left[T^e\mathbb{S}_Q(2\mathbf{y}-\mathbf{x})T^bT^a\mathbb{S}^T_Q(\mathbf{x})-T^e\mathbb{S}_Q(\mathbf{x})T^aT^b\mathbb{S}^T_Q(2\mathbf{y}-\mathbf{x})\right]\frac{\partial}{\partial \mathbb{S}^{eb}_Q(\mathbf{y})}\nonumber\\
&-&{\rm Tr}\left[T^e\mathbb{S}_Q(2\mathbf{x}-\mathbf{y})T^c\mathbb{S}^T_Q(\mathbf{y})\right]\frac{\partial}{\partial \mathbb{S}^{eb}_Q(\mathbf{x})}T^a_{bc}   \Bigg]\,.
\end{eqnarray}

We can easily check that this expression is UV finite. To see this we need to show that the sum of the integrands in the last two terms vanish for $\mathbf{y}\rightarrow\mathbf{x}$. 
This is obviously the case, since the $SU(N)$ generators in the penultimate line in this limit combine into $[T^a,T^b]$, and exactly cancel the term in the last line.


Now for the left charge
\begin{equation}
J^a_L(\mathbf x)=-{\rm Tr}\left[\frac{\partial}{\partial S(\mathbf{x})}S^T(\mathbf{x})T^a\right]\,.
\end{equation}
The same algebra as above yields
\begin{eqnarray}\label{jlq}
J^a_L(\mathbf x)&=&-{\rm Tr}\left[\frac{\partial}{\partial {\mathbb S}_Q(\mathbf{x})}{\mathbb S}^T_Q(\mathbf{x})T^a\right]+\frac{\alpha_s\beta_0^g} {4\pi^2N_c}\int^{\frac{1}{2Q}}_{\mathbf{y-x}}\frac{1}{(\mathbf{x}-\mathbf{y})^2}\nonumber\\
&\times&\Bigg[{\rm Tr}\left[T^aT^e\mathbb{S}_Q(2\mathbf{y}-\mathbf{x})T^b\mathbb{S}^T_Q(\mathbf{x})-T^eT^a\mathbb{S}_Q(\mathbf{x})T^b\mathbb{S}^T_Q(2\mathbf{y}-\mathbf{x})\right]\frac{\partial}{\partial \mathbb{S}^{eb}_Q(\mathbf{y})}\nonumber\\
&-&{\rm Tr}\left[T^c\mathbb{S}_Q(2\mathbf{x}-\mathbf{y}))T^b\mathbb{S}^T_Q(\mathbf{y})\right]\frac{\partial}{\partial \mathbb{S}^{eb}_Q(\mathbf{x})}T^a_{ce}   \Bigg]\,.
\end{eqnarray}

Again, it is easy to check that this expression is UV finite.

It is natural to define the " dressed color charge" operators that measure the charge of the dressed gluons: 
\begin{equation}
J^a_{Q,R}(\mathbf {x})=-{\rm Tr}\left[\frac{\partial}{\partial \mathbb{S}_Q(\mathbf{x})}T^a\mathbb{S}^T_Q(\mathbf{x})\right];\ \ \ \ \
J^a_{Q,L}(\mathbf{x})=-{\rm Tr}\left[\frac{\partial}{\partial {\mathbb S}_Q(\mathbf{x})}{\mathbb S}^T_Q(\mathbf{x})T^a\right]\,.
\end{equation}

As alluded to in the text, one can use these operators, rather than the original $J^a_{R(L)}$ to write the resummed form of $H_{\rm JIMWLK}$. Although this may look natural in a certain sense, it does not make any difference in perturbation theory. The fact that Eqs.\eqref{jrq} and \eqref{jlq} are free from the UV divergences means that the relation between the the ``bare charge'' operators and the ``dressed charge'' operators in purely perturbative and does not contain large logarithms. Thus the choice of which charge operators to use in the perturbatively resummed Hamiltonian is equivalent to the choice of the factorisation scheme. At the leading order, at which we work in this paper, the choice is immaterial. However if one would want to keep all $O(\alpha_s^2)$ terms, including those that are not enhanced by the transverse logarithms, this choice has to be explicitly specified.

\section{Appendix: Fourier transforming $\alpha_Q$}
To calculate $\alpha_Q(\mathbf{z})$ we need to calculate the following two dimensional Fourier transform:

\begin{align}
\label{Eq:Isx}
G_Q(x) &= 
\int \frac{d^2 p}{(2\pi)^2} e^{i \mathbf{p x}}
 \\ & \times
\left[\theta(4Q^2-p^2)+\left(\frac{p^2}{Q^2}\right)^{\frac{\alpha_s\beta_0}{4\pi}}\theta(Q_T^2-p^2)\theta(p^2-Q^2)+\left(\frac{Q_T^2}{Q^2}\right)^{\frac{\alpha_s\beta_0}{4\pi}}\theta(p^2-Q_T^2)\right]\,.\notag
\end{align}
We split the integral into three contributions corresponding to the order of terms in Eq.~\eqref{Eq:Isx}. 

The first integral can be analytically computed
\begin{align}
        I_1 = \int \frac{d^2 p}{(2\pi)^2} e^{i \mathbf{p x}}  \theta(Q^2-p^2) 
    =  \frac{Q J_1( Q x)}{\pi x}.  
\end{align}
Its asymptotics for small and large argument $Qx$  are
\begin{align}
    I_1(x\ll \frac{1}{Q}) &\approx \frac{Q^2}{\pi} , \\
    I_1(x\gg \frac{1}{Q}) &\approx -\frac{Q^{1/2}\cos \left( Q x+\frac{\pi }{4}\right)}{\pi^{3/2} x^{3/2}} \,.
\end{align}
The corrections to the latter are of order ${\cal O} (x^{-5/2})$.

The third integral reduces to delta function and the same analytic integral as in $I_1$
\begin{align}
        I_3 = \int \frac{d^2 p}{(2\pi)^2} e^{i \mathbf{p x}}  \left(\frac{Q_T^2}{Q^2}\right)^{\frac{\alpha_s\beta_0}{4\pi}}\theta(p^2-Q_T^2)
    = 
    \left(\frac{Q_T^2}{Q^2}\right)^{\frac{\alpha_s\beta_0} {4\pi}}\delta^{(2)}(\mathbf{x})
    - \left(\frac{Q_T^2}{Q^2}\right)^{\frac{\alpha_s\beta_0} {4\pi}}
 \frac{Q_T J_1(Q_T x)}{ \pi 
 x} \,.
\end{align}

The integral of the second term in Eq.~\eqref{Eq:Isx} leads to hypergeometric functions and is not very useful. Here we will only consider its assymptics. For small argument, 
\begin{align}
    I_2 (x \ll \frac1{ Q_T}) & \approx  
    \frac{Q^2}{\pi}
    \int_{1}^{(Q_T/Q)^2} dy   y^{\frac{\alpha_s \beta_0}{4 \pi}} = 
 \frac{Q^2}{\pi} \frac1{1 + \frac{\alpha_s \beta_0}{4 \pi}} \left( \left[\frac{Q_T^2}{Q^2} \right]^{1+ \frac{\alpha_s \beta_0}{4 \pi}}-1\right)
 \notag \\ &\approx 
  \frac{Q^2}{\pi}  \left( \left[\frac{Q_T^2}{Q^2} \right]^{1+ \frac{\alpha_s \beta_0}{4 \pi}}-1\right)\,.
\end{align}
For large argument $Q x\gg 1$, 
\begin{align}
  I_2(x \gg \frac{1}{Q}) \approx
     \frac{Q^{1/2}\cos \left( Q x+\frac{\pi }{4}\right)}{\pi^{3/2} x^{3/2}} 
     - 
     \left( \frac{Q_T^2}{Q^2} \right)^{\frac{\alpha_s \beta_0}{4 \pi}}
    \frac{Q_T^{1/2}\cos \left(Q_T x+\frac{\pi }{4}\right)}{\pi^{3/2} x^{3/2}} \, 
\end{align}
with  corrections of order  ${\cal O} (x^{-5/2})$. 
And finally, for the intermediate values, 
$\frac{1}{Q_T} \ll x \ll \frac{1}{Q_T}$, 
\begin{align}
  I_2(x \gg \frac{1}{Q}) \approx
     -\frac{Q^2}{\pi }
     -{\frac{\alpha_s \beta_0}{4 \pi}} \frac{ Q^2 }{\pi   (Q^2 x^2)^{1+{\frac{\alpha_s \beta_0}{4 \pi}}}}
     - 
     \left( \frac{Q_T^2}{Q^2} \right)^{\frac{\alpha_s \beta_0}{4 \pi}}
    \frac{Q_T^{1/2}\cos \left( Q_T x+\frac{\pi }{4}\right)}{\pi^{3/2} x^{3/2}} \,. 
\end{align}

From this set of results, one obtains $G(x)$ in different regions as presented in  Eqs.~\eqref{sx},~\eqref{ix}, and~\eqref{lx}.

\begin{acknowledgments}
We thank T. Altinoluk, I.~Balitsky, and G. Beuf for illuminating discussions. 
A.K. is supported by the NSF Nuclear Theory grant 2208387. This material is based
upon work supported by the U.S. Department of Energy, Office of Science, Office of Nuclear
Physics through the Contract No. DE-SC0020081 (V.S.) and the Saturated Glue (SURGE) Topical Collaboration. 
M.L. is supported by the US-Israel Binational Science Foundation grant \#2021789.
This work has been performed in the framework
of the MSCA RISE 823947 ``Heavy
ion collisions: collectivity and precision in saturation physics'' (HIEIC). V.S. thanks the ExtreMe Matter Institute for partial support and A. Andronic for hospitality at Physics Department of  Muenster University.

\end{acknowledgments}


\bibliography{apssamp}

\providecommand{\noopsort}[1]{}\providecommand{\singleletter}[1]{#1}%
\begin{thebibliography}{50}%
\makeatletter
\providecommand \@ifxundefined [1]{%
 \@ifx{#1\undefined}
}%
\providecommand \@ifnum [1]{%
 \ifnum #1\expandafter \@firstoftwo
 \else \expandafter \@secondoftwo
 \fi
}%
\providecommand \@ifx [1]{%
 \ifx #1\expandafter \@firstoftwo
 \else \expandafter \@secondoftwo
 \fi
}%
\providecommand \natexlab [1]{#1}%
\providecommand \enquote  [1]{``#1''}%
\providecommand \bibnamefont  [1]{#1}%
\providecommand \bibfnamefont [1]{#1}%
\providecommand \citenamefont [1]{#1}%
\providecommand \href@noop [0]{\@secondoftwo}%
\providecommand \href [0]{\begingroup \@sanitize@url \@href}%
\providecommand \@href[1]{\@@startlink{#1}\@@href}%
\providecommand \@@href[1]{\endgroup#1\@@endlink}%
\providecommand \@sanitize@url [0]{\catcode `\\12\catcode `\$12\catcode
  `\&12\catcode `\#12\catcode `\^12\catcode `\_12\catcode `\%12\relax}%
\providecommand \@@startlink[1]{}%
\providecommand \@@endlink[0]{}%
\providecommand \url  [0]{\begingroup\@sanitize@url \@url }%
\providecommand \@url [1]{\endgroup\@href {#1}{\urlprefix }}%
\providecommand \urlprefix  [0]{URL }%
\providecommand \Eprint [0]{\href }%
\providecommand \doibase [0]{https://doi.org/}%
\providecommand \selectlanguage [0]{\@gobble}%
\providecommand \bibinfo  [0]{\@secondoftwo}%
\providecommand \bibfield  [0]{\@secondoftwo}%
\providecommand \translation [1]{[#1]}%
\providecommand \BibitemOpen [0]{}%
\providecommand \bibitemStop [0]{}%
\providecommand \bibitemNoStop [0]{.\EOS\space}%
\providecommand \EOS [0]{\spacefactor3000\relax}%
\providecommand \BibitemShut  [1]{\csname bibitem#1\endcsname}%
\let\auto@bib@innerbib\@empty
\bibitem [{\citenamefont {Gelis}\ \emph {et~al.}(2010)\citenamefont {Gelis},
  \citenamefont {Iancu}, \citenamefont {Jalilian-Marian},\ and\ \citenamefont
  {Venugopalan}}]{Gelis:2010nm}%
  \BibitemOpen
  \bibfield  {author} {\bibinfo {author} {\bibfnamefont {F.}~\bibnamefont
  {Gelis}}, \bibinfo {author} {\bibfnamefont {E.}~\bibnamefont {Iancu}},
  \bibinfo {author} {\bibfnamefont {J.}~\bibnamefont {Jalilian-Marian}},\ and\
  \bibinfo {author} {\bibfnamefont {R.}~\bibnamefont {Venugopalan}},\
  }\bibfield  {title} {\bibinfo {title} {{The Color Glass Condensate}},\ }\href
  {https://doi.org/10.1146/annurev.nucl.010909.083629} {\bibfield  {journal}
  {\bibinfo  {journal} {Ann. Rev. Nucl. Part. Sci.}\ }\textbf {\bibinfo
  {volume} {60}},\ \bibinfo {pages} {463} (\bibinfo {year} {2010})},\ \Eprint
  {https://arxiv.org/abs/1002.0333} {arXiv:1002.0333 [hep-ph]} \BibitemShut
  {NoStop}%
\bibitem [{\citenamefont {Jalilian-Marian}\ \emph {et~al.}(1997)\citenamefont
  {Jalilian-Marian}, \citenamefont {Kovner}, \citenamefont {Leonidov},\ and\
  \citenamefont {Weigert}}]{Jalilian-Marian:1997qno}%
  \BibitemOpen
  \bibfield  {author} {\bibinfo {author} {\bibfnamefont {J.}~\bibnamefont
  {Jalilian-Marian}}, \bibinfo {author} {\bibfnamefont {A.}~\bibnamefont
  {Kovner}}, \bibinfo {author} {\bibfnamefont {A.}~\bibnamefont {Leonidov}},\
  and\ \bibinfo {author} {\bibfnamefont {H.}~\bibnamefont {Weigert}},\
  }\bibfield  {title} {\bibinfo {title} {{The BFKL equation from the Wilson
  renormalization group}},\ }\href
  {https://doi.org/10.1016/S0550-3213(97)00440-9} {\bibfield  {journal}
  {\bibinfo  {journal} {Nucl. Phys. B}\ }\textbf {\bibinfo {volume} {504}},\
  \bibinfo {pages} {415} (\bibinfo {year} {1997})},\ \Eprint
  {https://arxiv.org/abs/hep-ph/9701284} {arXiv:hep-ph/9701284} \BibitemShut
  {NoStop}%
\bibitem [{\citenamefont {Jalilian-Marian}\ \emph {et~al.}(1998)\citenamefont
  {Jalilian-Marian}, \citenamefont {Kovner},\ and\ \citenamefont
  {Weigert}}]{Jalilian-Marian:1997ubg}%
  \BibitemOpen
  \bibfield  {author} {\bibinfo {author} {\bibfnamefont {J.}~\bibnamefont
  {Jalilian-Marian}}, \bibinfo {author} {\bibfnamefont {A.}~\bibnamefont
  {Kovner}},\ and\ \bibinfo {author} {\bibfnamefont {H.}~\bibnamefont
  {Weigert}},\ }\bibfield  {title} {\bibinfo {title} {{The Wilson
  renormalization group for low x physics: Gluon evolution at finite parton
  density}},\ }\href {https://doi.org/10.1103/PhysRevD.59.014015} {\bibfield
  {journal} {\bibinfo  {journal} {Phys. Rev. D}\ }\textbf {\bibinfo {volume}
  {59}},\ \bibinfo {pages} {014015} (\bibinfo {year} {1998})},\ \Eprint
  {https://arxiv.org/abs/hep-ph/9709432} {arXiv:hep-ph/9709432} \BibitemShut
  {NoStop}%
\bibitem [{\citenamefont {Kovner}\ \emph {et~al.}(2000)\citenamefont {Kovner},
  \citenamefont {Milhano},\ and\ \citenamefont {Weigert}}]{Kovner:2000pt}%
  \BibitemOpen
  \bibfield  {author} {\bibinfo {author} {\bibfnamefont {A.}~\bibnamefont
  {Kovner}}, \bibinfo {author} {\bibfnamefont {J.~G.}\ \bibnamefont
  {Milhano}},\ and\ \bibinfo {author} {\bibfnamefont {H.}~\bibnamefont
  {Weigert}},\ }\bibfield  {title} {\bibinfo {title} {{Relating different
  approaches to nonlinear QCD evolution at finite gluon density}},\ }\href
  {https://doi.org/10.1103/PhysRevD.62.114005} {\bibfield  {journal} {\bibinfo
  {journal} {Phys. Rev. D}\ }\textbf {\bibinfo {volume} {62}},\ \bibinfo
  {pages} {114005} (\bibinfo {year} {2000})},\ \Eprint
  {https://arxiv.org/abs/hep-ph/0004014} {arXiv:hep-ph/0004014} \BibitemShut
  {NoStop}%
\bibitem [{\citenamefont {Iancu}\ \emph {et~al.}(2001)\citenamefont {Iancu},
  \citenamefont {Leonidov},\ and\ \citenamefont {McLerran}}]{Iancu:2000hn}%
  \BibitemOpen
  \bibfield  {author} {\bibinfo {author} {\bibfnamefont {E.}~\bibnamefont
  {Iancu}}, \bibinfo {author} {\bibfnamefont {A.}~\bibnamefont {Leonidov}},\
  and\ \bibinfo {author} {\bibfnamefont {L.~D.}\ \bibnamefont {McLerran}},\
  }\bibfield  {title} {\bibinfo {title} {{Nonlinear gluon evolution in the
  color glass condensate. 1.}},\ }\href
  {https://doi.org/10.1016/S0375-9474(01)00642-X} {\bibfield  {journal}
  {\bibinfo  {journal} {Nucl. Phys. A}\ }\textbf {\bibinfo {volume} {692}},\
  \bibinfo {pages} {583} (\bibinfo {year} {2001})},\ \Eprint
  {https://arxiv.org/abs/hep-ph/0011241} {arXiv:hep-ph/0011241} \BibitemShut
  {NoStop}%
\bibitem [{\citenamefont {Ferreiro}\ \emph {et~al.}(2002)\citenamefont
  {Ferreiro}, \citenamefont {Iancu}, \citenamefont {Leonidov},\ and\
  \citenamefont {McLerran}}]{Ferreiro:2001qy}%
  \BibitemOpen
  \bibfield  {author} {\bibinfo {author} {\bibfnamefont {E.}~\bibnamefont
  {Ferreiro}}, \bibinfo {author} {\bibfnamefont {E.}~\bibnamefont {Iancu}},
  \bibinfo {author} {\bibfnamefont {A.}~\bibnamefont {Leonidov}},\ and\
  \bibinfo {author} {\bibfnamefont {L.}~\bibnamefont {McLerran}},\ }\bibfield
  {title} {\bibinfo {title} {{Nonlinear gluon evolution in the color glass
  condensate. 2.}},\ }\href {https://doi.org/10.1016/S0375-9474(01)01329-X}
  {\bibfield  {journal} {\bibinfo  {journal} {Nucl. Phys. A}\ }\textbf
  {\bibinfo {volume} {703}},\ \bibinfo {pages} {489} (\bibinfo {year}
  {2002})},\ \Eprint {https://arxiv.org/abs/hep-ph/0109115}
  {arXiv:hep-ph/0109115} \BibitemShut {NoStop}%
\bibitem [{\citenamefont {Kovner}\ \emph
  {et~al.}(2014{\natexlab{a}})\citenamefont {Kovner}, \citenamefont
  {Lublinsky},\ and\ \citenamefont {Mulian}}]{Kovner:2013ona}%
  \BibitemOpen
  \bibfield  {author} {\bibinfo {author} {\bibfnamefont {A.}~\bibnamefont
  {Kovner}}, \bibinfo {author} {\bibfnamefont {M.}~\bibnamefont {Lublinsky}},\
  and\ \bibinfo {author} {\bibfnamefont {Y.}~\bibnamefont {Mulian}},\
  }\bibfield  {title} {\bibinfo {title} {{Jalilian-Marian, Iancu, McLerran,
  Weigert, Leonidov, Kovner evolution at next to leading order}},\ }\href
  {https://doi.org/10.1103/PhysRevD.89.061704} {\bibfield  {journal} {\bibinfo
  {journal} {Phys. Rev. D}\ }\textbf {\bibinfo {volume} {89}},\ \bibinfo
  {pages} {061704} (\bibinfo {year} {2014}{\natexlab{a}})},\ \Eprint
  {https://arxiv.org/abs/1310.0378} {arXiv:1310.0378 [hep-ph]} \BibitemShut
  {NoStop}%
\bibitem [{\citenamefont {Kovner}\ \emph
  {et~al.}(2014{\natexlab{b}})\citenamefont {Kovner}, \citenamefont
  {Lublinsky},\ and\ \citenamefont {Mulian}}]{Kovner:2014lca}%
  \BibitemOpen
  \bibfield  {author} {\bibinfo {author} {\bibfnamefont {A.}~\bibnamefont
  {Kovner}}, \bibinfo {author} {\bibfnamefont {M.}~\bibnamefont {Lublinsky}},\
  and\ \bibinfo {author} {\bibfnamefont {Y.}~\bibnamefont {Mulian}},\
  }\bibfield  {title} {\bibinfo {title} {{NLO JIMWLK evolution unabridged}},\
  }\href {https://doi.org/10.1007/JHEP08(2014)114} {\bibfield  {journal}
  {\bibinfo  {journal} {JHEP}\ }\textbf {\bibinfo {volume} {08}},\ \bibinfo
  {pages} {114}},\ \Eprint {https://arxiv.org/abs/1405.0418} {arXiv:1405.0418
  [hep-ph]} \BibitemShut {NoStop}%
\bibitem [{\citenamefont {Lublinsky}\ and\ \citenamefont
  {Mulian}(2017)}]{Lublinsky:2016meo}%
  \BibitemOpen
  \bibfield  {author} {\bibinfo {author} {\bibfnamefont {M.}~\bibnamefont
  {Lublinsky}}\ and\ \bibinfo {author} {\bibfnamefont {Y.}~\bibnamefont
  {Mulian}},\ }\bibfield  {title} {\bibinfo {title} {{High Energy QCD at NLO:
  from light-cone wave function to JIMWLK evolution}},\ }\href
  {https://doi.org/10.1007/JHEP05(2017)097} {\bibfield  {journal} {\bibinfo
  {journal} {JHEP}\ }\textbf {\bibinfo {volume} {05}},\ \bibinfo {pages}
  {097}},\ \Eprint {https://arxiv.org/abs/1610.03453} {arXiv:1610.03453
  [hep-ph]} \BibitemShut {NoStop}%
\bibitem [{\citenamefont {Balitsky}(1996)}]{Balitsky:1995ub}%
  \BibitemOpen
  \bibfield  {author} {\bibinfo {author} {\bibfnamefont {I.}~\bibnamefont
  {Balitsky}},\ }\bibfield  {title} {\bibinfo {title} {{Operator expansion for
  high-energy scattering}},\ }\href
  {https://doi.org/10.1016/0550-3213(95)00638-9} {\bibfield  {journal}
  {\bibinfo  {journal} {Nucl. Phys. B}\ }\textbf {\bibinfo {volume} {463}},\
  \bibinfo {pages} {99} (\bibinfo {year} {1996})},\ \Eprint
  {https://arxiv.org/abs/hep-ph/9509348} {arXiv:hep-ph/9509348} \BibitemShut
  {NoStop}%
\bibitem [{\citenamefont {Balitsky}(1997)}]{Balitsky:1997mk}%
  \BibitemOpen
  \bibfield  {author} {\bibinfo {author} {\bibfnamefont {I.}~\bibnamefont
  {Balitsky}},\ }\bibfield  {title} {\bibinfo {title} {{Operator expansion for
  diffractive high-energy scattering}},\ }\href
  {https://doi.org/10.1063/1.53693} {\bibfield  {journal} {\bibinfo  {journal}
  {AIP Conf. Proc.}\ }\textbf {\bibinfo {volume} {407}},\ \bibinfo {pages}
  {953} (\bibinfo {year} {1997})},\ \Eprint
  {https://arxiv.org/abs/hep-ph/9706411} {arXiv:hep-ph/9706411} \BibitemShut
  {NoStop}%
\bibitem [{\citenamefont {Kovchegov}(1999)}]{Kovchegov:1999yj}%
  \BibitemOpen
  \bibfield  {author} {\bibinfo {author} {\bibfnamefont {Y.~V.}\ \bibnamefont
  {Kovchegov}},\ }\bibfield  {title} {\bibinfo {title} {{Small x F(2) structure
  function of a nucleus including multiple pomeron exchanges}},\ }\href
  {https://doi.org/10.1103/PhysRevD.60.034008} {\bibfield  {journal} {\bibinfo
  {journal} {Phys. Rev. D}\ }\textbf {\bibinfo {volume} {60}},\ \bibinfo
  {pages} {034008} (\bibinfo {year} {1999})},\ \Eprint
  {https://arxiv.org/abs/hep-ph/9901281} {arXiv:hep-ph/9901281} \BibitemShut
  {NoStop}%
\bibitem [{\citenamefont {Balitsky}\ and\ \citenamefont
  {Chirilli}(2008)}]{Balitsky:2007feb}%
  \BibitemOpen
  \bibfield  {author} {\bibinfo {author} {\bibfnamefont {I.}~\bibnamefont
  {Balitsky}}\ and\ \bibinfo {author} {\bibfnamefont {G.~A.}\ \bibnamefont
  {Chirilli}},\ }\bibfield  {title} {\bibinfo {title} {{Next-to-leading order
  evolution of color dipoles}},\ }\href
  {https://doi.org/10.1103/PhysRevD.77.014019} {\bibfield  {journal} {\bibinfo
  {journal} {Phys. Rev. D}\ }\textbf {\bibinfo {volume} {77}},\ \bibinfo
  {pages} {014019} (\bibinfo {year} {2008})},\ \Eprint
  {https://arxiv.org/abs/0710.4330} {arXiv:0710.4330 [hep-ph]} \BibitemShut
  {NoStop}%
\bibitem [{\citenamefont {Kutak}\ and\ \citenamefont
  {Stasto}(2005)}]{Kutak:2004ym}%
  \BibitemOpen
  \bibfield  {author} {\bibinfo {author} {\bibfnamefont {K.}~\bibnamefont
  {Kutak}}\ and\ \bibinfo {author} {\bibfnamefont {A.~M.}\ \bibnamefont
  {Stasto}},\ }\bibfield  {title} {\bibinfo {title} {{Unintegrated gluon
  distribution from modified BK equation}},\ }\href
  {https://doi.org/10.1140/epjc/s2005-02223-0} {\bibfield  {journal} {\bibinfo
  {journal} {Eur. Phys. J. C}\ }\textbf {\bibinfo {volume} {41}},\ \bibinfo
  {pages} {343} (\bibinfo {year} {2005})},\ \Eprint
  {https://arxiv.org/abs/hep-ph/0408117} {arXiv:hep-ph/0408117} \BibitemShut
  {NoStop}%
\bibitem [{\citenamefont {Motyka}\ and\ \citenamefont
  {Stasto}(2009)}]{Motyka:2009gi}%
  \BibitemOpen
  \bibfield  {author} {\bibinfo {author} {\bibfnamefont {L.}~\bibnamefont
  {Motyka}}\ and\ \bibinfo {author} {\bibfnamefont {A.~M.}\ \bibnamefont
  {Stasto}},\ }\bibfield  {title} {\bibinfo {title} {{Exact kinematics in the
  small x evolution of the color dipole and gluon cascade}},\ }\href
  {https://doi.org/10.1103/PhysRevD.79.085016} {\bibfield  {journal} {\bibinfo
  {journal} {Phys. Rev. D}\ }\textbf {\bibinfo {volume} {79}},\ \bibinfo
  {pages} {085016} (\bibinfo {year} {2009})},\ \Eprint
  {https://arxiv.org/abs/0901.4949} {arXiv:0901.4949 [hep-ph]} \BibitemShut
  {NoStop}%
\bibitem [{\citenamefont {Beuf}(2014)}]{Beuf:2014uia}%
  \BibitemOpen
  \bibfield  {author} {\bibinfo {author} {\bibfnamefont {G.}~\bibnamefont
  {Beuf}},\ }\bibfield  {title} {\bibinfo {title} {{Improving the kinematics
  for low-$x$ QCD evolution equations in coordinate space}},\ }\href
  {https://doi.org/10.1103/PhysRevD.89.074039} {\bibfield  {journal} {\bibinfo
  {journal} {Phys. Rev. D}\ }\textbf {\bibinfo {volume} {89}},\ \bibinfo
  {pages} {074039} (\bibinfo {year} {2014})},\ \Eprint
  {https://arxiv.org/abs/1401.0313} {arXiv:1401.0313 [hep-ph]} \BibitemShut
  {NoStop}%
\bibitem [{\citenamefont {Iancu}\ \emph
  {et~al.}(2015{\natexlab{a}})\citenamefont {Iancu}, \citenamefont {Madrigal},
  \citenamefont {Mueller}, \citenamefont {Soyez},\ and\ \citenamefont
  {Triantafyllopoulos}}]{Iancu:2015vea}%
  \BibitemOpen
  \bibfield  {author} {\bibinfo {author} {\bibfnamefont {E.}~\bibnamefont
  {Iancu}}, \bibinfo {author} {\bibfnamefont {J.~D.}\ \bibnamefont {Madrigal}},
  \bibinfo {author} {\bibfnamefont {A.~H.}\ \bibnamefont {Mueller}}, \bibinfo
  {author} {\bibfnamefont {G.}~\bibnamefont {Soyez}},\ and\ \bibinfo {author}
  {\bibfnamefont {D.~N.}\ \bibnamefont {Triantafyllopoulos}},\ }\bibfield
  {title} {\bibinfo {title} {{Resumming double logarithms in the QCD evolution
  of color dipoles}},\ }\href {https://doi.org/10.1016/j.physletb.2015.03.068}
  {\bibfield  {journal} {\bibinfo  {journal} {Phys. Lett. B}\ }\textbf
  {\bibinfo {volume} {744}},\ \bibinfo {pages} {293} (\bibinfo {year}
  {2015}{\natexlab{a}})},\ \Eprint {https://arxiv.org/abs/1502.05642}
  {arXiv:1502.05642 [hep-ph]} \BibitemShut {NoStop}%
\bibitem [{\citenamefont {Iancu}\ \emph
  {et~al.}(2015{\natexlab{b}})\citenamefont {Iancu}, \citenamefont {Madrigal},
  \citenamefont {Mueller}, \citenamefont {Soyez},\ and\ \citenamefont
  {Triantafyllopoulos}}]{Iancu:2015joa}%
  \BibitemOpen
  \bibfield  {author} {\bibinfo {author} {\bibfnamefont {E.}~\bibnamefont
  {Iancu}}, \bibinfo {author} {\bibfnamefont {J.~D.}\ \bibnamefont {Madrigal}},
  \bibinfo {author} {\bibfnamefont {A.~H.}\ \bibnamefont {Mueller}}, \bibinfo
  {author} {\bibfnamefont {G.}~\bibnamefont {Soyez}},\ and\ \bibinfo {author}
  {\bibfnamefont {D.~N.}\ \bibnamefont {Triantafyllopoulos}},\ }\bibfield
  {title} {\bibinfo {title} {{Collinearly-improved BK evolution meets the HERA
  data}},\ }\href {https://doi.org/10.1016/j.physletb.2015.09.071} {\bibfield
  {journal} {\bibinfo  {journal} {Phys. Lett. B}\ }\textbf {\bibinfo {volume}
  {750}},\ \bibinfo {pages} {643} (\bibinfo {year} {2015}{\natexlab{b}})},\
  \Eprint {https://arxiv.org/abs/1507.03651} {arXiv:1507.03651 [hep-ph]}
  \BibitemShut {NoStop}%
\bibitem [{\citenamefont {Sabio~Vera}(2005)}]{Vera:2005jt}%
  \BibitemOpen
  \bibfield  {author} {\bibinfo {author} {\bibfnamefont {A.}~\bibnamefont
  {Sabio~Vera}},\ }\bibfield  {title} {\bibinfo {title} {{An 'All-poles'
  approximation to collinear resummations in the Regge limit of perturbative
  QCD}},\ }\href {https://doi.org/10.1016/j.nuclphysb.2005.06.003} {\bibfield
  {journal} {\bibinfo  {journal} {Nucl. Phys. B}\ }\textbf {\bibinfo {volume}
  {722}},\ \bibinfo {pages} {65} (\bibinfo {year} {2005})},\ \Eprint
  {https://arxiv.org/abs/hep-ph/0505128} {arXiv:hep-ph/0505128} \BibitemShut
  {NoStop}%
\bibitem [{\citenamefont {Duclou\'e}\ \emph {et~al.}(2019)\citenamefont
  {Duclou\'e}, \citenamefont {Iancu}, \citenamefont {Mueller}, \citenamefont
  {Soyez},\ and\ \citenamefont {Triantafyllopoulos}}]{Ducloue:2019ezk}%
  \BibitemOpen
  \bibfield  {author} {\bibinfo {author} {\bibfnamefont {B.}~\bibnamefont
  {Duclou\'e}}, \bibinfo {author} {\bibfnamefont {E.}~\bibnamefont {Iancu}},
  \bibinfo {author} {\bibfnamefont {A.~H.}\ \bibnamefont {Mueller}}, \bibinfo
  {author} {\bibfnamefont {G.}~\bibnamefont {Soyez}},\ and\ \bibinfo {author}
  {\bibfnamefont {D.~N.}\ \bibnamefont {Triantafyllopoulos}},\ }\bibfield
  {title} {\bibinfo {title} {{Non-linear evolution in QCD at high-energy beyond
  leading order}},\ }\href {https://doi.org/10.1007/JHEP04(2019)081} {\bibfield
   {journal} {\bibinfo  {journal} {JHEP}\ }\textbf {\bibinfo {volume} {04}},\
  \bibinfo {pages} {081}},\ \Eprint {https://arxiv.org/abs/1902.06637}
  {arXiv:1902.06637 [hep-ph]} \BibitemShut {NoStop}%
\bibitem [{\citenamefont {Balitsky}(2007)}]{Balitsky:2006wa}%
  \BibitemOpen
  \bibfield  {author} {\bibinfo {author} {\bibfnamefont {I.}~\bibnamefont
  {Balitsky}},\ }\bibfield  {title} {\bibinfo {title} {{Quark contribution to
  the small-x evolution of color dipole}},\ }\href
  {https://doi.org/10.1103/PhysRevD.75.014001} {\bibfield  {journal} {\bibinfo
  {journal} {Phys. Rev. D}\ }\textbf {\bibinfo {volume} {75}},\ \bibinfo
  {pages} {014001} (\bibinfo {year} {2007})},\ \Eprint
  {https://arxiv.org/abs/hep-ph/0609105} {arXiv:hep-ph/0609105} \BibitemShut
  {NoStop}%
\bibitem [{\citenamefont {Kovchegov}\ and\ \citenamefont
  {Weigert}(2007)}]{Kovchegov:2006vj}%
  \BibitemOpen
  \bibfield  {author} {\bibinfo {author} {\bibfnamefont {Y.~V.}\ \bibnamefont
  {Kovchegov}}\ and\ \bibinfo {author} {\bibfnamefont {H.}~\bibnamefont
  {Weigert}},\ }\bibfield  {title} {\bibinfo {title} {{Triumvirate of Running
  Couplings in Small-x Evolution}},\ }\href
  {https://doi.org/10.1016/j.nuclphysa.2006.10.075} {\bibfield  {journal}
  {\bibinfo  {journal} {Nucl. Phys. A}\ }\textbf {\bibinfo {volume} {784}},\
  \bibinfo {pages} {188} (\bibinfo {year} {2007})},\ \Eprint
  {https://arxiv.org/abs/hep-ph/0609090} {arXiv:hep-ph/0609090} \BibitemShut
  {NoStop}%
\bibitem [{\citenamefont {Gardi}\ \emph {et~al.}(2007)\citenamefont {Gardi},
  \citenamefont {Kuokkanen}, \citenamefont {Rummukainen},\ and\ \citenamefont
  {Weigert}}]{Gardi:2006rp}%
  \BibitemOpen
  \bibfield  {author} {\bibinfo {author} {\bibfnamefont {E.}~\bibnamefont
  {Gardi}}, \bibinfo {author} {\bibfnamefont {J.}~\bibnamefont {Kuokkanen}},
  \bibinfo {author} {\bibfnamefont {K.}~\bibnamefont {Rummukainen}},\ and\
  \bibinfo {author} {\bibfnamefont {H.}~\bibnamefont {Weigert}},\ }\bibfield
  {title} {\bibinfo {title} {{Running coupling and power corrections in
  nonlinear evolution at the high-energy limit}},\ }\href
  {https://doi.org/10.1016/j.nuclphysa.2006.12.004} {\bibfield  {journal}
  {\bibinfo  {journal} {Nucl. Phys. A}\ }\textbf {\bibinfo {volume} {784}},\
  \bibinfo {pages} {282} (\bibinfo {year} {2007})},\ \Eprint
  {https://arxiv.org/abs/hep-ph/0609087} {arXiv:hep-ph/0609087} \BibitemShut
  {NoStop}%
\bibitem [{\citenamefont {Altinoluk}\ \emph {et~al.}(2023)\citenamefont
  {Altinoluk}, \citenamefont {Beuf}, \citenamefont {Lublinsky},\ and\
  \citenamefont {Skokov}}]{Altinoluk:2023}%
  \BibitemOpen
  \bibfield  {author} {\bibinfo {author} {\bibfnamefont {T.}~\bibnamefont
  {Altinoluk}}, \bibinfo {author} {\bibfnamefont {G.}~\bibnamefont {Beuf}},
  \bibinfo {author} {\bibfnamefont {M.}~\bibnamefont {Lublinsky}},\ and\
  \bibinfo {author} {\bibfnamefont {V.}~\bibnamefont {Skokov}},\ }\bibfield
  {title} {\bibinfo {title} {{Negativity of running coupling prescriptions in
  JIMWLK}},\ }\href@noop {} {\  (\bibinfo {year} {2023})},\ \Eprint
  {https://arxiv.org/abs/to be published} {to be published} \BibitemShut
  {NoStop}%
\bibitem [{\citenamefont {Dokshitzer}(1977)}]{Dokshitzer:1977sg}%
  \BibitemOpen
  \bibfield  {author} {\bibinfo {author} {\bibfnamefont {Y.~L.}\ \bibnamefont
  {Dokshitzer}},\ }\bibfield  {title} {\bibinfo {title} {{Calculation of the
  Structure Functions for Deep Inelastic Scattering and e+ e- Annihilation by
  Perturbation Theory in Quantum Chromodynamics.}},\ }\href@noop {} {\bibfield
  {journal} {\bibinfo  {journal} {Sov. Phys. JETP}\ }\textbf {\bibinfo {volume}
  {46}},\ \bibinfo {pages} {641} (\bibinfo {year} {1977})}\BibitemShut
  {NoStop}%
\bibitem [{\citenamefont {Gribov}\ and\ \citenamefont
  {Lipatov}(1972)}]{Gribov:1972ri}%
  \BibitemOpen
  \bibfield  {author} {\bibinfo {author} {\bibfnamefont {V.~N.}\ \bibnamefont
  {Gribov}}\ and\ \bibinfo {author} {\bibfnamefont {L.~N.}\ \bibnamefont
  {Lipatov}},\ }\bibfield  {title} {\bibinfo {title} {{Deep inelastic e p
  scattering in perturbation theory}},\ }\href@noop {} {\bibfield  {journal}
  {\bibinfo  {journal} {Sov. J. Nucl. Phys.}\ }\textbf {\bibinfo {volume}
  {15}},\ \bibinfo {pages} {438} (\bibinfo {year} {1972})}\BibitemShut
  {NoStop}%
\bibitem [{\citenamefont {Altarelli}\ and\ \citenamefont
  {Parisi}(1977)}]{Altarelli:1977zs}%
  \BibitemOpen
  \bibfield  {author} {\bibinfo {author} {\bibfnamefont {G.}~\bibnamefont
  {Altarelli}}\ and\ \bibinfo {author} {\bibfnamefont {G.}~\bibnamefont
  {Parisi}},\ }\bibfield  {title} {\bibinfo {title} {{Asymptotic Freedom in
  Parton Language}},\ }\href {https://doi.org/10.1016/0550-3213(77)90384-4}
  {\bibfield  {journal} {\bibinfo  {journal} {Nucl. Phys. B}\ }\textbf
  {\bibinfo {volume} {126}},\ \bibinfo {pages} {298} (\bibinfo {year}
  {1977})}\BibitemShut {NoStop}%
\bibitem [{\citenamefont {Kovner}(2005)}]{Kovner:2005pe}%
  \BibitemOpen
  \bibfield  {author} {\bibinfo {author} {\bibfnamefont {A.}~\bibnamefont
  {Kovner}},\ }\bibfield  {title} {\bibinfo {title} {{High energy evolution:
  The Wave function point of view}},\ }\href@noop {} {\bibfield  {journal}
  {\bibinfo  {journal} {Acta Phys. Polon. B}\ }\textbf {\bibinfo {volume}
  {36}},\ \bibinfo {pages} {3551} (\bibinfo {year} {2005})},\ \Eprint
  {https://arxiv.org/abs/hep-ph/0508232} {arXiv:hep-ph/0508232} \BibitemShut
  {NoStop}%
\bibitem [{\citenamefont {Kovchegov}\ and\ \citenamefont
  {Levin}(2013)}]{Kovchegov:2012mbw}%
  \BibitemOpen
  \bibfield  {author} {\bibinfo {author} {\bibfnamefont {Y.~V.}\ \bibnamefont
  {Kovchegov}}\ and\ \bibinfo {author} {\bibfnamefont {E.}~\bibnamefont
  {Levin}},\ }\href {https://doi.org/10.1017/9781009291446} {\emph {\bibinfo
  {title} {{Quantum Chromodynamics at High Energy}}}},\ Vol.~\bibinfo {volume}
  {33}\ (\bibinfo  {publisher} {Oxford University Press},\ \bibinfo {year}
  {2013})\BibitemShut {NoStop}%
\bibitem [{\citenamefont {Braun}(1995)}]{Braun:1994mw}%
  \BibitemOpen
  \bibfield  {author} {\bibinfo {author} {\bibfnamefont {M.~A.}\ \bibnamefont
  {Braun}},\ }\bibfield  {title} {\bibinfo {title} {{Reggeized gluons with a
  running coupling constant}},\ }\href
  {https://doi.org/10.1016/0370-2693(95)00101-P} {\bibfield  {journal}
  {\bibinfo  {journal} {Phys. Lett. B}\ }\textbf {\bibinfo {volume} {348}},\
  \bibinfo {pages} {190} (\bibinfo {year} {1995})},\ \Eprint
  {https://arxiv.org/abs/hep-ph/9408261} {arXiv:hep-ph/9408261} \BibitemShut
  {NoStop}%
\bibitem [{\citenamefont {Levin}(1995)}]{Levin:1994di}%
  \BibitemOpen
  \bibfield  {author} {\bibinfo {author} {\bibfnamefont {E.}~\bibnamefont
  {Levin}},\ }\bibfield  {title} {\bibinfo {title} {{Renormalons at low x}},\
  }\href {https://doi.org/10.1016/0550-3213(95)00416-P} {\bibfield  {journal}
  {\bibinfo  {journal} {Nucl. Phys. B}\ }\textbf {\bibinfo {volume} {453}},\
  \bibinfo {pages} {303} (\bibinfo {year} {1995})},\ \Eprint
  {https://arxiv.org/abs/hep-ph/9412345} {arXiv:hep-ph/9412345} \BibitemShut
  {NoStop}%
\bibitem [{\citenamefont {Chirilli}\ and\ \citenamefont
  {Kovchegov}(2013)}]{Chirilli:2013kca}%
  \BibitemOpen
  \bibfield  {author} {\bibinfo {author} {\bibfnamefont {G.~A.}\ \bibnamefont
  {Chirilli}}\ and\ \bibinfo {author} {\bibfnamefont {Y.~V.}\ \bibnamefont
  {Kovchegov}},\ }\bibfield  {title} {\bibinfo {title} {{Solution of the NLO
  BFKL Equation and a Strategy for Solving the All-Order BFKL Equation}},\
  }\href {https://doi.org/10.1007/JHEP06(2013)055} {\bibfield  {journal}
  {\bibinfo  {journal} {JHEP}\ }\textbf {\bibinfo {volume} {06}},\ \bibinfo
  {pages} {055}},\ \Eprint {https://arxiv.org/abs/1305.1924} {arXiv:1305.1924
  [hep-ph]} \BibitemShut {NoStop}%
\bibitem [{\citenamefont {Duan}\ \emph {et~al.}(2023)\citenamefont {Duan},
  \citenamefont {Skokov},\ and\ \citenamefont {Stephens}}]{Duan}%
  \BibitemOpen
  \bibfield  {author} {\bibinfo {author} {\bibfnamefont {H.}~\bibnamefont
  {Duan}}, \bibinfo {author} {\bibfnamefont {V.}~\bibnamefont {Skokov}},\ and\
  \bibinfo {author} {\bibfnamefont {C.}~\bibnamefont {Stephens}},\ }\bibfield
  {title} {\bibinfo {title} {{Color neutralization scale at small x}},\
  }\href@noop {} {\  (\bibinfo {year} {2023})},\ \Eprint
  {https://arxiv.org/abs/to be published} {to be published} \BibitemShut
  {NoStop}%
\bibitem [{\citenamefont {Salam}(1998)}]{Salam:1998tj}%
  \BibitemOpen
  \bibfield  {author} {\bibinfo {author} {\bibfnamefont {G.~P.}\ \bibnamefont
  {Salam}},\ }\bibfield  {title} {\bibinfo {title} {{A Resummation of large
  subleading corrections at small x}},\ }\href
  {https://doi.org/10.1088/1126-6708/1998/07/019} {\bibfield  {journal}
  {\bibinfo  {journal} {JHEP}\ }\textbf {\bibinfo {volume} {07}},\ \bibinfo
  {pages} {019}},\ \Eprint {https://arxiv.org/abs/hep-ph/9806482}
  {arXiv:hep-ph/9806482} \BibitemShut {NoStop}%
\bibitem [{\citenamefont {Ciafaloni}\ \emph
  {et~al.}(1999{\natexlab{a}})\citenamefont {Ciafaloni}, \citenamefont
  {Colferai},\ and\ \citenamefont {Salam}}]{Ciafaloni:1999au}%
  \BibitemOpen
  \bibfield  {author} {\bibinfo {author} {\bibfnamefont {M.}~\bibnamefont
  {Ciafaloni}}, \bibinfo {author} {\bibfnamefont {D.}~\bibnamefont
  {Colferai}},\ and\ \bibinfo {author} {\bibfnamefont {G.~P.}\ \bibnamefont
  {Salam}},\ }\bibfield  {title} {\bibinfo {title} {{A collinear model for
  small x physics}},\ }\href {https://doi.org/10.1088/1126-6708/1999/10/017}
  {\bibfield  {journal} {\bibinfo  {journal} {JHEP}\ }\textbf {\bibinfo
  {volume} {10}},\ \bibinfo {pages} {017}},\ \Eprint
  {https://arxiv.org/abs/hep-ph/9907409} {arXiv:hep-ph/9907409} \BibitemShut
  {NoStop}%
\bibitem [{\citenamefont {Ciafaloni}\ \emph
  {et~al.}(1999{\natexlab{b}})\citenamefont {Ciafaloni}, \citenamefont
  {Colferai},\ and\ \citenamefont {Salam}}]{Ciafaloni:1999yw}%
  \BibitemOpen
  \bibfield  {author} {\bibinfo {author} {\bibfnamefont {M.}~\bibnamefont
  {Ciafaloni}}, \bibinfo {author} {\bibfnamefont {D.}~\bibnamefont
  {Colferai}},\ and\ \bibinfo {author} {\bibfnamefont {G.~P.}\ \bibnamefont
  {Salam}},\ }\bibfield  {title} {\bibinfo {title} {{Renormalization group
  improved small x equation}},\ }\href
  {https://doi.org/10.1103/PhysRevD.60.114036} {\bibfield  {journal} {\bibinfo
  {journal} {Phys. Rev. D}\ }\textbf {\bibinfo {volume} {60}},\ \bibinfo
  {pages} {114036} (\bibinfo {year} {1999}{\natexlab{b}})},\ \Eprint
  {https://arxiv.org/abs/hep-ph/9905566} {arXiv:hep-ph/9905566} \BibitemShut
  {NoStop}%
\bibitem [{\citenamefont {Ciafaloni}\ \emph
  {et~al.}(2003{\natexlab{a}})\citenamefont {Ciafaloni}, \citenamefont
  {Colferai}, \citenamefont {Colferai}, \citenamefont {Salam},\ and\
  \citenamefont {Stasto}}]{Ciafaloni:2003ek}%
  \BibitemOpen
  \bibfield  {author} {\bibinfo {author} {\bibfnamefont {M.}~\bibnamefont
  {Ciafaloni}}, \bibinfo {author} {\bibfnamefont {D.}~\bibnamefont {Colferai}},
  \bibinfo {author} {\bibfnamefont {D.}~\bibnamefont {Colferai}}, \bibinfo
  {author} {\bibfnamefont {G.~P.}\ \bibnamefont {Salam}},\ and\ \bibinfo
  {author} {\bibfnamefont {A.~M.}\ \bibnamefont {Stasto}},\ }\bibfield  {title}
  {\bibinfo {title} {{Extending QCD perturbation theory to higher energies}},\
  }\href {https://doi.org/10.1016/j.physletb.2003.09.078} {\bibfield  {journal}
  {\bibinfo  {journal} {Phys. Lett. B}\ }\textbf {\bibinfo {volume} {576}},\
  \bibinfo {pages} {143} (\bibinfo {year} {2003}{\natexlab{a}})},\ \Eprint
  {https://arxiv.org/abs/hep-ph/0305254} {arXiv:hep-ph/0305254} \BibitemShut
  {NoStop}%
\bibitem [{\citenamefont {Ciafaloni}\ \emph
  {et~al.}(2003{\natexlab{b}})\citenamefont {Ciafaloni}, \citenamefont
  {Colferai}, \citenamefont {Salam},\ and\ \citenamefont
  {Stasto}}]{Ciafaloni:2003rd}%
  \BibitemOpen
  \bibfield  {author} {\bibinfo {author} {\bibfnamefont {M.}~\bibnamefont
  {Ciafaloni}}, \bibinfo {author} {\bibfnamefont {D.}~\bibnamefont {Colferai}},
  \bibinfo {author} {\bibfnamefont {G.~P.}\ \bibnamefont {Salam}},\ and\
  \bibinfo {author} {\bibfnamefont {A.~M.}\ \bibnamefont {Stasto}},\ }\bibfield
   {title} {\bibinfo {title} {{Renormalization group improved small x Green's
  function}},\ }\href {https://doi.org/10.1103/PhysRevD.68.114003} {\bibfield
  {journal} {\bibinfo  {journal} {Phys. Rev. D}\ }\textbf {\bibinfo {volume}
  {68}},\ \bibinfo {pages} {114003} (\bibinfo {year} {2003}{\natexlab{b}})},\
  \Eprint {https://arxiv.org/abs/hep-ph/0307188} {arXiv:hep-ph/0307188}
  \BibitemShut {NoStop}%
\bibitem [{\citenamefont {Ciafaloni}\ \emph {et~al.}(2004)\citenamefont
  {Ciafaloni}, \citenamefont {Colferai}, \citenamefont {Salam},\ and\
  \citenamefont {Stasto}}]{Ciafaloni:2003kd}%
  \BibitemOpen
  \bibfield  {author} {\bibinfo {author} {\bibfnamefont {M.}~\bibnamefont
  {Ciafaloni}}, \bibinfo {author} {\bibfnamefont {D.}~\bibnamefont {Colferai}},
  \bibinfo {author} {\bibfnamefont {G.~P.}\ \bibnamefont {Salam}},\ and\
  \bibinfo {author} {\bibfnamefont {A.~M.}\ \bibnamefont {Stasto}},\ }\bibfield
   {title} {\bibinfo {title} {{The Gluon splitting function at moderately small
  x}},\ }\href {https://doi.org/10.1016/j.physletb.2004.02.054} {\bibfield
  {journal} {\bibinfo  {journal} {Phys. Lett. B}\ }\textbf {\bibinfo {volume}
  {587}},\ \bibinfo {pages} {87} (\bibinfo {year} {2004})},\ \Eprint
  {https://arxiv.org/abs/hep-ph/0311325} {arXiv:hep-ph/0311325} \BibitemShut
  {NoStop}%
\bibitem [{\citenamefont {Thorne}(2001)}]{Thorne:2001nr}%
  \BibitemOpen
  \bibfield  {author} {\bibinfo {author} {\bibfnamefont {R.~S.}\ \bibnamefont
  {Thorne}},\ }\bibfield  {title} {\bibinfo {title} {{The Running coupling BFKL
  anomalous dimensions and splitting functions}},\ }\href
  {https://doi.org/10.1103/PhysRevD.64.074005} {\bibfield  {journal} {\bibinfo
  {journal} {Phys. Rev. D}\ }\textbf {\bibinfo {volume} {64}},\ \bibinfo
  {pages} {074005} (\bibinfo {year} {2001})},\ \Eprint
  {https://arxiv.org/abs/hep-ph/0103210} {arXiv:hep-ph/0103210} \BibitemShut
  {NoStop}%
\bibitem [{\citenamefont {White}\ and\ \citenamefont
  {Thorne}(2007)}]{White:2006yh}%
  \BibitemOpen
  \bibfield  {author} {\bibinfo {author} {\bibfnamefont {C.~D.}\ \bibnamefont
  {White}}\ and\ \bibinfo {author} {\bibfnamefont {R.~S.}\ \bibnamefont
  {Thorne}},\ }\bibfield  {title} {\bibinfo {title} {{A Global Fit to
  Scattering Data with NLL BFKL Resummations}},\ }\href
  {https://doi.org/10.1103/PhysRevD.75.034005} {\bibfield  {journal} {\bibinfo
  {journal} {Phys. Rev. D}\ }\textbf {\bibinfo {volume} {75}},\ \bibinfo
  {pages} {034005} (\bibinfo {year} {2007})},\ \Eprint
  {https://arxiv.org/abs/hep-ph/0611204} {arXiv:hep-ph/0611204} \BibitemShut
  {NoStop}%
\bibitem [{\citenamefont {Altarelli}\ \emph {et~al.}(2000)\citenamefont
  {Altarelli}, \citenamefont {Ball},\ and\ \citenamefont
  {Forte}}]{Altarelli:1999vw}%
  \BibitemOpen
  \bibfield  {author} {\bibinfo {author} {\bibfnamefont {G.}~\bibnamefont
  {Altarelli}}, \bibinfo {author} {\bibfnamefont {R.~D.}\ \bibnamefont
  {Ball}},\ and\ \bibinfo {author} {\bibfnamefont {S.}~\bibnamefont {Forte}},\
  }\bibfield  {title} {\bibinfo {title} {{Resummation of singlet parton
  evolution at small x}},\ }\href
  {https://doi.org/10.1016/S0550-3213(00)00032-8} {\bibfield  {journal}
  {\bibinfo  {journal} {Nucl. Phys. B}\ }\textbf {\bibinfo {volume} {575}},\
  \bibinfo {pages} {313} (\bibinfo {year} {2000})},\ \Eprint
  {https://arxiv.org/abs/hep-ph/9911273} {arXiv:hep-ph/9911273} \BibitemShut
  {NoStop}%
\bibitem [{\citenamefont {Altarelli}\ \emph {et~al.}(2001)\citenamefont
  {Altarelli}, \citenamefont {Ball},\ and\ \citenamefont
  {Forte}}]{Altarelli:2000mh}%
  \BibitemOpen
  \bibfield  {author} {\bibinfo {author} {\bibfnamefont {G.}~\bibnamefont
  {Altarelli}}, \bibinfo {author} {\bibfnamefont {R.~D.}\ \bibnamefont
  {Ball}},\ and\ \bibinfo {author} {\bibfnamefont {S.}~\bibnamefont {Forte}},\
  }\bibfield  {title} {\bibinfo {title} {{Small x resummation and HERA
  structure function data}},\ }\href
  {https://doi.org/10.1016/S0550-3213(01)00023-2} {\bibfield  {journal}
  {\bibinfo  {journal} {Nucl. Phys. B}\ }\textbf {\bibinfo {volume} {599}},\
  \bibinfo {pages} {383} (\bibinfo {year} {2001})},\ \Eprint
  {https://arxiv.org/abs/hep-ph/0011270} {arXiv:hep-ph/0011270} \BibitemShut
  {NoStop}%
\bibitem [{\citenamefont {Altarelli}\ \emph {et~al.}(2002)\citenamefont
  {Altarelli}, \citenamefont {Ball},\ and\ \citenamefont
  {Forte}}]{Altarelli:2001ji}%
  \BibitemOpen
  \bibfield  {author} {\bibinfo {author} {\bibfnamefont {G.}~\bibnamefont
  {Altarelli}}, \bibinfo {author} {\bibfnamefont {R.~D.}\ \bibnamefont
  {Ball}},\ and\ \bibinfo {author} {\bibfnamefont {S.}~\bibnamefont {Forte}},\
  }\bibfield  {title} {\bibinfo {title} {{Factorization and resummation of
  small x scaling violations with running coupling}},\ }\href
  {https://doi.org/10.1016/S0550-3213(01)00563-6} {\bibfield  {journal}
  {\bibinfo  {journal} {Nucl. Phys. B}\ }\textbf {\bibinfo {volume} {621}},\
  \bibinfo {pages} {359} (\bibinfo {year} {2002})},\ \Eprint
  {https://arxiv.org/abs/hep-ph/0109178} {arXiv:hep-ph/0109178} \BibitemShut
  {NoStop}%
\bibitem [{\citenamefont {Altarelli}\ \emph {et~al.}(2003)\citenamefont
  {Altarelli}, \citenamefont {Ball},\ and\ \citenamefont
  {Forte}}]{Altarelli:2003hk}%
  \BibitemOpen
  \bibfield  {author} {\bibinfo {author} {\bibfnamefont {G.}~\bibnamefont
  {Altarelli}}, \bibinfo {author} {\bibfnamefont {R.~D.}\ \bibnamefont
  {Ball}},\ and\ \bibinfo {author} {\bibfnamefont {S.}~\bibnamefont {Forte}},\
  }\bibfield  {title} {\bibinfo {title} {{An Anomalous dimension for small x
  evolution}},\ }\href {https://doi.org/10.1016/j.nuclphysb.2003.09.040}
  {\bibfield  {journal} {\bibinfo  {journal} {Nucl. Phys. B}\ }\textbf
  {\bibinfo {volume} {674}},\ \bibinfo {pages} {459} (\bibinfo {year}
  {2003})},\ \Eprint {https://arxiv.org/abs/hep-ph/0306156}
  {arXiv:hep-ph/0306156} \BibitemShut {NoStop}%
\bibitem [{\citenamefont {Altarelli}\ \emph {et~al.}(2008)\citenamefont
  {Altarelli}, \citenamefont {Ball},\ and\ \citenamefont
  {Forte}}]{Altarelli:2008aj}%
  \BibitemOpen
  \bibfield  {author} {\bibinfo {author} {\bibfnamefont {G.}~\bibnamefont
  {Altarelli}}, \bibinfo {author} {\bibfnamefont {R.~D.}\ \bibnamefont
  {Ball}},\ and\ \bibinfo {author} {\bibfnamefont {S.}~\bibnamefont {Forte}},\
  }\bibfield  {title} {\bibinfo {title} {{Small x Resummation with Quarks:
  Deep-Inelastic Scattering}},\ }\href
  {https://doi.org/10.1016/j.nuclphysb.2008.03.003} {\bibfield  {journal}
  {\bibinfo  {journal} {Nucl. Phys. B}\ }\textbf {\bibinfo {volume} {799}},\
  \bibinfo {pages} {199} (\bibinfo {year} {2008})},\ \Eprint
  {https://arxiv.org/abs/0802.0032} {arXiv:0802.0032 [hep-ph]} \BibitemShut
  {NoStop}%
\bibitem [{\citenamefont {Ball}\ \emph {et~al.}(2018)\citenamefont {Ball},
  \citenamefont {Bertone}, \citenamefont {Bonvini}, \citenamefont {Marzani},
  \citenamefont {Rojo},\ and\ \citenamefont {Rottoli}}]{Ball:2017otu}%
  \BibitemOpen
  \bibfield  {author} {\bibinfo {author} {\bibfnamefont {R.~D.}\ \bibnamefont
  {Ball}}, \bibinfo {author} {\bibfnamefont {V.}~\bibnamefont {Bertone}},
  \bibinfo {author} {\bibfnamefont {M.}~\bibnamefont {Bonvini}}, \bibinfo
  {author} {\bibfnamefont {S.}~\bibnamefont {Marzani}}, \bibinfo {author}
  {\bibfnamefont {J.}~\bibnamefont {Rojo}},\ and\ \bibinfo {author}
  {\bibfnamefont {L.}~\bibnamefont {Rottoli}},\ }\bibfield  {title} {\bibinfo
  {title} {{Parton distributions with small-x resummation: evidence for BFKL
  dynamics in HERA data}},\ }\href
  {https://doi.org/10.1140/epjc/s10052-018-5774-4} {\bibfield  {journal}
  {\bibinfo  {journal} {Eur. Phys. J. C}\ }\textbf {\bibinfo {volume} {78}},\
  \bibinfo {pages} {321} (\bibinfo {year} {2018})},\ \Eprint
  {https://arxiv.org/abs/1710.05935} {arXiv:1710.05935 [hep-ph]} \BibitemShut
  {NoStop}%
\bibitem [{\citenamefont {Jalilian-Marian}(2019)}]{Jalilian-Marian:2018iui}%
  \BibitemOpen
  \bibfield  {author} {\bibinfo {author} {\bibfnamefont {J.}~\bibnamefont
  {Jalilian-Marian}},\ }\bibfield  {title} {\bibinfo {title} {{Quark jets
  scattering from a gluon field: from saturation to high $p_t$}},\ }\href
  {https://doi.org/10.1103/PhysRevD.99.014043} {\bibfield  {journal} {\bibinfo
  {journal} {Phys. Rev. D}\ }\textbf {\bibinfo {volume} {99}},\ \bibinfo
  {pages} {014043} (\bibinfo {year} {2019})},\ \Eprint
  {https://arxiv.org/abs/1809.04625} {arXiv:1809.04625 [hep-ph]} \BibitemShut
  {NoStop}%
\bibitem [{\citenamefont {Boussarie}\ and\ \citenamefont
  {Mehtar-Tani}(2022{\natexlab{a}})}]{Boussarie:2020fpb}%
  \BibitemOpen
  \bibfield  {author} {\bibinfo {author} {\bibfnamefont {R.}~\bibnamefont
  {Boussarie}}\ and\ \bibinfo {author} {\bibfnamefont {Y.}~\bibnamefont
  {Mehtar-Tani}},\ }\bibfield  {title} {\bibinfo {title} {{A novel formulation
  of the unintegrated gluon distribution for DIS}},\ }\href
  {https://doi.org/10.1016/j.physletb.2022.137125} {\bibfield  {journal}
  {\bibinfo  {journal} {Phys. Lett. B}\ }\textbf {\bibinfo {volume} {831}},\
  \bibinfo {pages} {137125} (\bibinfo {year} {2022}{\natexlab{a}})},\ \Eprint
  {https://arxiv.org/abs/2006.14569} {arXiv:2006.14569 [hep-ph]} \BibitemShut
  {NoStop}%
\bibitem [{\citenamefont {Boussarie}\ and\ \citenamefont
  {Mehtar-Tani}(2022{\natexlab{b}})}]{Boussarie:2021wkn}%
  \BibitemOpen
  \bibfield  {author} {\bibinfo {author} {\bibfnamefont {R.}~\bibnamefont
  {Boussarie}}\ and\ \bibinfo {author} {\bibfnamefont {Y.}~\bibnamefont
  {Mehtar-Tani}},\ }\bibfield  {title} {\bibinfo {title} {{Gluon-mediated
  inclusive Deep Inelastic Scattering from Regge to Bjorken kinematics}},\
  }\href {https://doi.org/10.1007/JHEP07(2022)080} {\bibfield  {journal}
  {\bibinfo  {journal} {JHEP}\ }\textbf {\bibinfo {volume} {07}},\ \bibinfo
  {pages} {080}},\ \Eprint {https://arxiv.org/abs/2112.01412} {arXiv:2112.01412
  [hep-ph]} \BibitemShut {NoStop}%
\end{thebibliography}%

\end{document}